# STRATEGIC STALEMATES: THE PARADOX OF EXPORT CONTROLS IN THE U.S.-CHINA AI RACE

JINGWEN LIU* AND JYH-AN LEE**

ABSTRACT

Export control functions as both a policy tool and legislative instrument, crafted to safeguard national interests by regulating the export of specific goods and technologies to targeted foreign nations. This mechanism has become central in the intensifying technological and geopolitical rivalry between the United States and China, particularly in sensitive sectors such as artificial intelligence (AI). The scope of these controls encompasses a broad range of items—from advanced computing chips and capital to personnel and critical minerals vital to semiconductor manufacturing.

Since October 2022, the U.S. Department of Commerce's Bureau of Industry and Security (BIS) has progressively tightened restrictions on the export of certain advanced computing components and integrated circuit applications to China. These measures have been intensified through subsequent actions in October 2023, April 2024, December 2024, and January 2025. In response, China has enacted its own export restrictions on critical minerals and rare earth elements essential to semiconductor production. Additionally, China has lodged a complaint with the World Trade Organization (WTO), accusing the United States of violating several provisions of the General Agreement on Tariffs and Trade (GATT).

This Article contends that, although export controls serve as strategic instruments in the U.S.-China AI competition, their long-term effectiveness is questionable. Far from stifling technological development, such measures often inadvertently promote self-reliance and the growth of independent research and development. Furthermore, the adoption of stringent and, at times arbitrary,

* Research Fellow, Centre for Legal Innovation and Digital Society (CLINDS), The Chinese University of Hong Kong Faculty of Law.

** Professor and Executive Director, Centre for Legal Innovation and Digital Society (CLINDS), The Chinese University of Hong Kong Faculty of Law.

*An earlier draft of this paper was presented at the 17th Intellectual Property Conference: AI Beyond Imagination at the Chinese University of Hong Kong and the 2025 GILJ International Law Symposium on BRICS Plus and the Evolving Technological and Financial World Order at Gonzaga Law School. We are grateful for the valuable insights and constructive feedback provided by Upendra Acharya, Albert Wai-Kit Chan, Dongmin Chen, George K. Foster, Joseph Isanga, Daryl Lim, Wai Shun Lo, Peter Yu, and Dongsheng Zang. This research was supported by the CUHK Direct Grant for Research (Grant #4059076).*

export restrictions can lead to violations of WTO obligations, complicating dispute resolution processes and impeding overall progress in AI technology. Furthermore, this study explores the legal implications of the overuse of export controls under international law. By advocating for a more restrained interpretation of security interests, this Article argues that AI models and semiconductors intended for commercial or dual-use purposes do not meet the criteria for security exceptions outlined in Article XXI(b) of GATT.

## I. INTRODUCTION

Artificial intelligence (AI) has fundamentally transformed the global economy and redefined the structure of international leadership. While nearly all nations recognize the strategic importance of their domestic AI research and development (R&D), the technological rivalry between the United States and China—two leaders in this race—has arguably emerged as the most significant aspect of global AI competition. In this context, both countries have implemented policies designed not only to promote the growth of their own AI sector but also to impede the progress of their rival. Among the measures employed, export controls have been frequently utilized by both nations amidst the intensifying AI race.

Starting from October 2022, the United States Department of Commerce's Bureau of Industry and Security (BIS) has been tightening control over the export of certain advanced computing chips, semiconductor chip manufacturing equipment, supercomputers and integrated circuit (IC) applications to China.[1] These measures were further escalated in October 2023,[2] April 2024,[3] December 2024[4] and January 2025.[5] In all these initiatives, the BIS attempted to justify the measures by resorting to U.S. national security, foreign policy interests and U.S. technological supremacy, specifically referencing to China's strategic goal of becoming a global AI leader by the year 2030.[6] Fighting back these measures, China introduced in retaliation a series of export restrictions on certain critical minerals and rare earth elements (REEs) essential to the production of semiconductor chips.[7] In

---

1. Implementation of Additional Export Controls: Certain Advanced Computing and Semiconductor Manufacturing Items; Supercomputer and Semiconductor End Use; Entity List Modification, 87 Fed. Reg. 62186 (Oct. 13, 2022) [hereinafter Implementation of Additional Export Controls October 2022].

2. Implementation of Additional Export Controls: Certain Advanced Computing Items; Supercomputer and Semiconductor End Use; Updates and Corrections (AC/S IFR), 88 Fed. Reg. 73458 (Oct. 25, 2023) [hereinafter AC/S IFR October 2023]; Export Controls on Semiconductor Manufacturing Items (SME IFR), 88 Fed. Reg. 73424 (Oct. 25, 2023) [hereinafter SME IFR October 2023]; Entity List Additions, 88 Fed. Reg. 71991 (Oct. 19, 2023) [hereinafter Entity List Additions October 2023].

3. Implementation of Additional Export Controls: Certain Advanced Computing Items; Supercomputer and Semiconductor End Use; Updates and Corrections; and Export Controls on Semiconductor Manufacturing Items; Corrections and Clarifications, 89 Fed. Reg. 23878 (Apr. 4, 2024) [hereinafter Implementation of Additional Export Controls April 2024]

4. Foreign-Produced Direct Product Rule Additions, and Refinements to Controls for Advanced Computing and Semiconductor Manufacturing Items, 89 Fed. Reg. 96790 (Dec. 5, 2024) [hereinafter Foreign-Produced Direct Product Rule Additions December 2024]; Additions and Modifications to the Entity List; Removals from the Validated End-User (VEU) Program, 89 Fed. Reg. 96830 (Dec. 5, 2024) [hereinafter Additions and Modifications to the Entity List December 2024].

5. Framework for Artificial Intelligence Diffusion, 90 Fed. Reg. 4544 (Jan. 15, 2025) [herein after Framework for Artificial Intelligence Diffusion January 2025].

6. Implementation of Additional Export Controls October 2022, *supra* note 1.

7. Guanyu Dui Jia, Zhe Xiangguan Wuxiang Shishi Chukou Guanzhi De Gonggao (关于对镓、锗相关物项实施出口管制的公告) [Notice on Implementing Export Control on Gallium- and Germanium-Related Substance], Notice of the Gen. Admin. of Customs No. 23 [2023] (promulgated by the General Admin. of Customs of the Ministry of Com., July 3, 2023, effective Aug. 1, 2023) [hereinafter Notice on Gallium and Germanium July 2023]; Guanyu Youhua Tiaozheng Shimo Wuxiang Linshi Chukou Guanzhi Cuoshi De Gonggao (关于优化调整石墨物项临时出口管制措施的公告) [Notice on Improving and Adjusting the Provisional Export Control Measures on Graphite], Notice of the Gen. Admin. of Customs No. 39 [2023] (promulgated by the Gen. Admin. of Customs of the Ministry of Com., Oct. 20, 2023, effective Dec. 1, 2023) [hereinafter Notice on Graphite October 2023]; Guanyu Dui Ti Deng Xiangguan addition, China also filed a complaint with the World Trade Organization (WTO) against the United States, alleging several violations of the General Agreement on Tariffs and Trade (GATT), among others.[8]

Despite the increasingly restrictive regulations and their subsequent implementation by both sides, the question of whether these measures will genuinely impede competitors' AI innovation and secure a nation's technological supremacy remains a subject of debate. For example, in response to export restrictions, China has been making substantial investments in its domestic chipmaking industry,[9] while the United States has focused on developing its domestic REE supply chains.[10] The launch of DeepSeek in 2025 by a company based in Hangzhou serves as a potential illustration of how these restrictions have failed to achieve their intended policy objectives, particularly in light of the stringent controls by the United States on access to advanced semiconductor chips.[11]

On the other hand, these measures have led to a significant rise in international trade disputes, as evidenced by the ongoing WTO consultation proceeding between the two nations. While countries justify their actions under the security exceptions stipulated by Article XXI of the GATT, citing the integration of AI technologies into military applications, it is important to note that the traditional understanding of "security" in international law is largely defense-

---

Wuxiang Shishi Chukou Guanzhi De Gonggao (关于对锑等相关物项实施出口管制的公告) [Notice on Implementing Export Control on Antimony and Other Related Substance], Notice of the Gen. Admin. of Customs No. 33 [2024] (promulgated by the Gen. Admin. of Customs of the Ministry of Com., Aug. 15, 2024, effective Sept. 15, 2024) [hereinafter Notice on Antimony August 2024]; Guanyu Jiaqiang Xiangguan Liangyong Wuzhi Dui Meiguo Chukou Guanzhi De Gonggao (关于加强相关两用物质对美国出口管制的公告) [Notice on Strengthening Export Control of Dual-Use Substance to the United States], Notice of the Gen. Admin. of Customs No. 46 [2024] (promulgated by the Gen, Admin. of Customs of the Ministry of Com., Dec. 3, 2024, effective Dec. 3, 2024) [hereinafter Notice on Dual-Use Substance December 2024]; Guanyu Dui Bufen Zhong Zhong Xitu Xiangguan Wuxiang Shishi Chukou Guanzhi De Jueding (关于对部分中重稀土相关物项实施出口管制的决定) [Notice on Implementing Export Control on Certain Heavy Rare Earth Substance], Notice of the Gen. Admin. of Customs No. 18 [2025] (promulgated by the Gen. Admin. of Customs of the Ministry of Com., Apr. 4, 2025, effective Apr. 4, 2025) [hereinafter Notice on Heavy Rare Earths April 2025].

8. United States — Measures on Certain Semiconductor and other Products, and Related Services and Technologies, WT/DS615 (complaint filed Dec. 12, 2022).

9. *Zhuce Ziben 3440 Yi, Guojia Dajijin Sanqi Laile! Huo Jiang Touzi Zhexie Zhongdian Xiangmu (*注册资本3440亿，国家大基金三期来了！或将投资这些重点项目*)* [*With 344 Billion Yuan Registered Capital, Here Comes Phase Three of the Big Fund! It May Invest in These Key Projects*], CCTV (May 28, 2025), https://finance.cctv.com/2024/05/28/ARTIhM6N831KrZyhRYMQziH8240528.shtml.

10. C. Todd Lopez, DOD Looks to Establish "Mine-to-Magnet" Supply Chain for Rare Earth Materials, DEPARTMENT OF DEFENSE (Mar. 11, 2024), https://www.defense.gov/News/News-Stories/Article/Article/3700059/dod-looks-to-establish-mine-to-magnet-supply-chain-for-rare-earth-materials/.

11. However, there are allegations that DeepSeek obtained Nvidia chips subject to export restrictions through intermediaries such as Singapore, which are denied by both Singapore and Nvidia. *See* Section III.1.B.

focused.[12] This perspective precludes commercial-use or dual-use AI components from being adequately protected under this exception.

In light of the above, this Article posits that, although export controls have been wielded as strategic tools in the AI competition between the United States and China,[13] they frequently fail to achieve the anticipated outcomes in the long term, inadvertently fostering self-reliance and independent R&D. Furthermore, by implementing stringent and often arbitrary export control measures, these nations risk violating their WTO obligations, thereby complicating dispute resolution and hindering the overall advancement of AI technology.

Section II of this Article introduces the export restrictions imposed by both parties upon their respective adversary. Section III examines the reasons why these restrictions may have proven less effective than anticipated, highlighting both countries' capacity to diversify their supply chains in the short term and to bolster independent R&D in the long run. Section IV then explores the international law implications of the weaponization of export restrictions. By advocating a more constrained interpretation of security interests, this Article argues that AI models and semiconductors intended for commercial or dual-use purposes do not meet the criteria for security exceptions under Article XXI(b) of the GATT.

## II. EXPORT RESTRICTIONS TARGETING THE AI SECTOR

### *A. The United States*

Export restrictions in the United States have encompassed AI models, semiconductors essential to the development of these models, as well as American investments in Chinese AI firms.

#### 1. AI Models

In May 2024, a proposed Enhancing National Frameworks for Overseas Restriction of Critical Exports (ENFORCE) Act was introduced in the United States House of Representatives, aimed at amending the existing Export Control Reform Act (ECRA) of 2018.[14]

---

12. Stephen Kho & Yujin K. McNamara, *Focus on China: The Expansive Use of National Security Measures to Address Economic Competitiveness Concerns*, 17 U. PA. ASIAN L. REV 368, 372 (2022).

13. Anu Bradford, Eileen Li & Matthew C. Waxman, *How Domestic Institutions Shape the Global Tech War*, 16 HARV NAT'L SEC J 75, 84 (2025).

14. Enhancing National Frameworks for Overseas Restriction of Critical Exports Act (ENFORCE Act), H.R. 8315, 118th Cong. (2024).

Under the current framework, the BIS lacks the authority to impose controls directly on AI models; instead, it can only regulate advanced semiconductor chips, which are classified as "dual-use items" (i.e., items with both military and commercial applications), to administer transactions with China in the AI sector.[15] Consequently, the new ENFORCE Act proposes a licensing scheme that would empower the BIS to require licenses for U.S. firms wishing to sell covered AI models to Chinese companies.[16] During the full committee markup of the Act, the House Foreign Affairs Committee Chairman Michael McCaul remarked

> AI has created a technology revolution that will determine whether America remains the world's leading superpower, or whether it gets eclipsed by China…We must understand that whoever sets the rules on its application will win this great power competition and determine the global balance of power… Our top AI companies could inadvertently fuel China's technological ascent, empowering their military and malign ambition. As the CCP [Chinese Communist Party] looks to expand their technological advancements to enhance their surveillance state and war machine, it is critical that we protect our sensitive technology from falling into their hands.[17]

The bill garnered an overwhelming majority vote in the Committee and still requires approval from the House, the Senate, and execution of the President before it can come into effect. If passed, it is expected that the array of export control measures currently implemented by the BIS, which primarily focus on the semiconductor sector, will rapidly expand to encompass downstream AI models, algorithms and parameters. This would elevate the existing level of foreign trade regulations and impact a broader range of companies, including OpenAI, Google and Microsoft.

---

15. H. R. FOREIGN AFF. COMM., ENHANCING NATIONAL FRAMEWORKS FOR OVERSEAS RESTRICTION OF CRITICAL EXPORTS (ENFORCE) ACT (2024).

16. *Id.*

17. Press Release, H. R. Foreign Aff. Comm., Chairman McCaul's ENFORCE Act Passes Out of Committee with Broad Bipartisan Support 43-3 (May 22, 2024), https://foreignaffairs.house.gov/news/press-releases/chairman-mccaul-s-enforce-act-passes-out-committee-with-broad-bipartisan-support-43-3.

2. Semiconductor Chips

As the BIS currently lacks authority to impose restrictions directly on the export of AI models, most existing measures aimed at curbing China's AI development focus on upstream semiconductors. The initial batch of restrictions was announced on October 7, 2022, instituting stringent export controls to China through a rigorous licensing regime targeting certain advanced computing semiconductor chips, transactions for supercomputer end uses, specific semiconductor manufacturing items, transactions for particular IC end uses and dealings involving certain entities on the Entity List.[18]

In its notice, the BIS justified its action by referencing its authority under the ECRA to regulate the export of commodities, software, technology, and activities for national security and foreign policy reasons. It stated that the prohibited items are used by China to produce advanced military systems including *weapons of mass destruction*; improve the speed and accuracy of its *military decision making, planning, and logistics*, as well as of its *autonomous military systems*; and commit human rights abuses.[19]

The measures were updated by the BIS on October 17, 2023, imposing more stringent licensing requirements and expanding the scope of items covered by the ban.[20] Revising three previous rules, this October 2023 update adjusted the parameters that determine whether a computing chip requires a license, enabling it to cover some less advanced chips. Furthermore, the update introduced controls on a wider array of semiconductor manufacturing equipment and enlarged the list of destinations requiring a license, aiming to mitigate the risk of Chinese entities bypassing existing restrictions via intermediaries.[21]

The ramifications of this augmented prohibition are palpable among technology companies in both the United States and China. For example, while the prior constraints were limited to only the most advanced chips, such as Nvidia's A100 and H100 data center chips, Nvidia had crafted alternative models tailored for their Chinese clientele, including the A800 and H800 GPUs, to navigate the October 2022 ban. Nevertheless, these modified models have

---

18. Implementation of Additional Export Controls October 2022, *supra* note 1.
19. *Id.*
20. AC/S IFR October 2023, *supra* note 2; SME IFR October 2023, *supra* note 2; Entity List Additions October 2023, *supra* note 2.
21. *Id.*

also fallen under the restrictions delineated in the October 2023 update.[22]

On April 4, 2024, the BIS further refined its licensing scheme with an update that adjusted certain license exceptions and restored controls of Export Control Classification Numbers (ECCNs) that incorporate .z paragraphs by removing the exceptions for .z paragraphs from the clauses pertaining to national security, missile technology, nuclear proliferation, and/or crime control license requirement paragraphs.[23] The BIS stated that these adjustments were implemented to address the potential risks of circumvention, such as embedding a chip within a .z item to qualify for a license exception.[24]

Subsequently, on December 2, 2024, the BIS added another 24 types of semiconductor manufacturing equipment and three types of software tools and high-bandwidth memory to the export control list. Moreover, 140 Chinese companies, including tool manufacturers, semiconductor fabs, and investment companies engaged in China's military modernization, were added to the Entity List.[25] In this suite of measures, the BIS continued to justify its actions on the grounds of national security, highlighting threats emanating from China's "*military-civil fusion strategy*," which supplements traditional military modernization efforts. Therefore, the objectives of the updated controls include not only decelerating China's AI integration into military applications but also impairing China's "development of an *indigenous semiconductor ecosystem*. . . ."[26]

On January 13, 2025, a mere week before the conclusion of the Biden-Harris Administration's term, the BIS released a new Framework for Artificial Intelligence Diffusion. This framework built upon earlier regulations and, while expanding existing controls on advanced computing ICs, introduced new controls on AI model weights for certain advanced dual-use AI models with closed-weight.[27] AI model weights, defined as "numerical parameter[s] within an AI model" play a pivotal role in determining the model's outputs in response to inputs,[28] thereby critically influencing an AI model's learning capabilities and the performance of sophisticated AI systems.[29] Despite continuing to assert its authority under the ECRA to safeguard national security and foreign policy interests, the BIS also acknowledged that "advanced AI models both pose unique threats to U.S. national security and foreign policy and have the potential to unlock *unique economic and social benefits*."[30]

---

22. Che Pan & Ann Cao, *How the Latest US Chip Export Controls Exposed China's Weak Link in the Semiconductor Supply Chain*, S. CHINA MORNING POST (Nov. 24, 2023), https://www.scmp.com/tech/tech-war/article/3239803/how-latest-us-chip-export-controls-exposed-chinas-weak-link-semiconductor-supply-chain.

23. Implementation of Additional Export Controls April 2024, *supra* note 3.

24. *Id.*

25. Foreign-Produced Direct Product Rule Additions December 2024, *supra* note 4; Additions and Modifications to the Entity List December 2024, *supra* note 4.

26. Press Release, U.S. Bureau of Indus. & Sec., Commerce Strengthens Export Controls to Restrict China's Capability to Produce Advanced Semiconductors for Military Applications (Dec. 2, 2024), https://media.bis.gov/sites/default/files/documents/FINAL%20DOC%20Nat%20Sec%20Action%20Rls%20Dec%202%2024.pdf.

27. Framework for Artificial Intelligence Diffusion January 2025, *supra* note 5.

On April 9, 2025, Nvidia revealed that the U.S. government had notified it of the inclusion of its H20 IC in the export control.[31] The H20, akin to the previously modified H800 and A800, is a less sophisticated variant derived from Nvidia's flagship H100 model. It has been specifically adapted for Chinese customers to navigate around the constraints of export controls. Allegations have surfaced suggesting that H20 chips have significantly contributed to the early 2025 success of DeepSeek.[32] It was not until July 16, 2025—one month after the London trade talks between high-level officials from both countries—that the ban on H20 chips finally lifted.[33] Interestingly, merely two weeks after Nvidia persuaded Washington to lift the export restrictions, the Cyberspace Administration of China summoned Nvidia representatives and demanded they clarify and submit relevant supporting documentation regarding security risks, including potential vulnerabilities and backdoors, associated with its H20 computing chips sold to China. [34] This move signaled a potential import ban, should Nvidia fail to meet these requirements. Notably, the tracking

28. Exec. Order No. 14110, 88 Fed. Reg. 75191 (Oct. 30, 2023).

29. Framework for Artificial Intelligence Diffusion: A Step Forward for US Security and Economic Strength in the Age of AI, CLIFFORD CHANCE (Feb. 2025), https://www.cliffordchance.com/content/dam/cliffordchance/briefings/2025/02/Framework%20for%20Artificial%20Intelligence%20Diffusion%20-%20A%20Step%20Forward%20for%20US%20Security%20and%20Economic%20Strength%20in%20the%20Age%20of%20AI.pdf.

30. Framework for Artificial Intelligence Diffusion January 2025, *supra* note 5, 4545.

31. Nvidia Corp., Current Report Pursuant to Section 13 or 15(d) of the Security Exchange Act of 1934, SEC. & EXCH. COMM'N (Apr. 9, 2025), https://www.sec.gov/ix?doc=/Archives/edgar/data/0001045810/000104581025000082/nvda-20250409.htm.

32. *See* Section III.1.B.

33. *Jensen Huang: Will Commence Sales of H20 Chips to the Chinese Market*, CCTV (July 15, 2025), https://content-static.cctvnews.cctv.com/snow-book/index.html?item_id=10051782561450980212&t=1752546250770&toc_style_id=feeds_default&share_to=wechat&track_id=cd320e5b-a206-4b19-8914-c52414acb09e.

34. *Guojia Hulianwang Xinxi Bangongshi Jiu H20 Suanli Xinpian Loudong Houmen Anquan Fengxian Yuetan Yingweida Gongsi (*国家互联网信息办公室就H20算力芯片漏洞后门安全风险约谈英伟达公司*)* [*Cyberspace Administration of China Summoned Nvidia Corp. regarding Security Risks Associated with H20 Computing Chips' Backdoor Vulnerabilities*], CYBERSPACE ADMIN. OF CHINA (July 31, 2025), https://www.cac.gov.cn/2025-07/31/c_1755675743897163.htm.

features cited by the Chinese regulators were themselves conditions imposed on Nvidia by U.S. authorities as part of the approval process of exporting H20.[35] This unexpected turn of events underscores the delicate and precarious nature of international trade in sensitive technologies amid contemporary geopolitical uncertainty.

Meanwhile, liability for violations of export controls ranges from civil penalties to criminal prosecution. In November 2024, the BIS, as part of a settlement deal, imposed a civil penalty of USD 500,000 on a New York-based semiconductor wafer manufacturer for shipping wafers valued at USD 17.1 million to a Chinese company listed on the Entity List without the requisite license.[36] In February 2025, a North Carolina businessman pleaded guilty to attempting to export accelerometer technology with military applications to China, having falsely labeled the end user as located in Missouri. The case remains under investigation, and the defendant faces a maximum sentence of 20 years in prison.[37]

From 2022 to 2025, the export restrictions on semiconductor chips have exhibited a trend of expansion across two dimensions. Geographically, the focus of export control has evolved from China in 2022, to encompassing a group of nations in a tiered system which progressively expanded during 2023 and 2024, culminating in a worldwide license requirement for some of the most advanced technologies by 2025. This tiered approach aligns the complexity of licensing requirements with the prevailing global alliance and rivalry structure, aiming to mitigate the risk of China obtaining American chips via third countries, thereby undermining "the national security interests of the United States and its allies."[38]

Concurrently, the scope of items subject to export control has also broadened. Initially, with the enactment of the October 2022 ban, the BIS voiced security concerns regarding China's incorporation of advanced AI systems, using American chips, into military applications and military decision-making processes.[39] However, in the subsequent years, the focus expanded from purely military to include dual-use items, with an increasing number of private-sector stakeholders appearing on the Entity List. Citing China's "military-civil fusion strategy," the BIS explicitly and repeatedly stated that the objectives of these stringent measures include not only addressing the security threats posed by China but also impeding the development of China's indigenous semiconductor ecosystem, thereby securing economic advantages.[40] As will be analyzed in Section IV below, this shift in focus carries profound implications for the trade dispute between the parties in terms of their obligations under international law.

---

35. *Id.*

36. Press Release, U.S. Bureau of Indus. & Sec., BIS Imposes $500,000 Mitigated Penalty Against GlobalFoundries For 74 Shipments to Entity Listed Chinese Firm (Nov. 1, 2025), https://media.bis.gov/sites/default/files/documents/Global%20Foundries%20Press%20Release_final.pdf.

37. Press Release, U.S. Dep't of Just. Off. of Pub. Aff., North Carolina Man Pleads Guilty to Attempting to Illegally Export Sensitive Technology to China (Feb. 28, 2025), https://www.justice.gov/opa/pr/north-carolina-man-pleads-guilty-attempting-illegally-export-sensitive-technology-china.

38. AC/S IFR October 2023, *supra* note 2.

39. Framework for Artificial Intelligence Diffusion January 2025, *supra* note 5, at 4546-48.

3. Capital and Personnel

In addition to semiconductors, the U.S. government also imposed restrictions on the export of American capital. In August 2023, the Biden Administration promulgated an executive order that barred outbound investments into China's AI sector.[41] This measure was justified by the White House on the grounds that, although the United States is generally an advocate for international investment, such activities must be in accord with national security interests of the country. Given that certain investments hold the potential to accelerate the development of sensitive technologies in nations that may use them to weaken the capabilities of the United States, they present an "unusual and extraordinary threat to the national security of the United States," thus constituting a national emergency that necessitates immediate action.[42] Subsequent to the proclamation of this order, it was reported that foreign investment in China's semiconductor sector saw a marked decline in the year of 2023.[43]

To implement this executive order, the U.S. Department of the Treasury, on October 28, 2024, promulgated a Final Rule that prohibits U.S. persons from partaking in specified transactions with Chinese persons (including those based in Hong Kong and Macau) involving three categories of technologies or products: semiconductors and microelectronics, quantum information technologies, and AI.[44] The Assistant Secretary for Investment Security, Paul Rosen, elucidated the necessity of these measures by stating that U.S. investments, including intangible advantages such as managerial expertise and access to networks of investment and talent, must not facilitate the development of "military, intelligence, and cyber capabilities" in concerning countries.[45]

---

40. *Id.*
41. Exec. Order No. 14105, 88 Fed. Reg. 54867 (Aug. 9, 2023).
42. *Id.*
43. Chang Che & John Liu, *"De-Americanize": How China Is Remaking Its Chip Business*, N.Y. TIMES (May 11, 2023), https://www.nytimes.com/2023/05/11/technology/china-us-chip-controls.html.
44. Provisions Pertaining to U.S. Investments in Certain National Security Technologies and Products in Countries of Concern, 89 Fed. Reg. 90398 (Nov. 15, 2024) [hereinafter Provisions Pertaining to U.S. Investments November 2024].

The Final Rule delineates the obligations of U.S. persons under this regulation, detailing the specifications of the technologies or products whose transactions are prescribed, and those which necessitate notification to the Treasury. It provides for certain exemptions, including investments in publicly traded securities, investments through index funds, mutual funds or exchange traded funds, intracompany transactions, buyouts of ownership, employment compensation in the form of stock or stock options, certain limited partnership investments, certain syndicated debt financings and investments made pursuant to pre-Outbound Order binding commitments.[46] Furthermore, the Rule specified the penalties for infractions, ranging from a fine of either USD 250,000 or twice the amount of the transaction, whichever is greater, to criminal liabilities including a fine of up to USD 1 million, or, imprisonment for up to 20 years for natural persons, or both.[47]

### *B. China*

In response to this series of restrictions, China implemented three types of measures: (1) diversifying its supply chain and reducing reliance on foreign resources (see Section III below); (2) seeking consultations at the WTO (see Section IV below); and (3) imposing retaliatory export controls targeting U.S. AI capabilities.

China's retaliatory actions include export controls on specific raw materials crucial for the production of high-frequency computer chips. On July 3, 2023, the Chinese Ministry of Commerce announced export controls on critical minerals such as gallium and germanium, which took effect on August 1, 2023.[48] Gallium, a soft and silvery metal, is widely used in semiconductor materials, optoelectronic devices, cancer and malaria chemotherapy, antimicrobial, and dental materials.[49] Germanium, on the other hand, is frequently used in high-tech applications, including infrared systems, fiber optics, polymer catalysis, electronics, and solar cells.[50] The control measures employed a licensing scheme akin to that imposed by the U.S. BIS. Specifically, any party wishing to export these materials must apply to the Ministry of Commerce through its provincial department for a license, providing clarification and proof of the end use of the exported goods and the identity of end users.[51] The Ministry of Commerce will decide on whether to approve an application, and, where necessary—such as when the exported goods have a "significant impact" on national security—it must seek further approval from the State Council.[52] Violations of these export controls could result in administrative and criminal liabilities.[53]

---

45. Press Release, U.S. Dep't of the Treasury, Treasury Issues Regulations to Implement Executive Order Addressing U.S. Investments in Certain National Security Technologies and Products in Countries of Concern (October 28, 2024), https://home.treasury.gov/news/press-releases/jy2687.

46. Provisions Pertaining to U.S. Investments November 2024, *supra* note 44, at 90400.

47. *Id.* at 90449.

48. Notice on Gallium and Germanium July 2023, *supra* note 7.

49. Akiyo Tanaka, Nikki Maples-Reynolds & Bruce A. Fowler, Gallium and Gallium Semiconductor Compounds, in HANDBOOK ON THE TOXICOLOGY OF METALS 275, 275 (Gunnar F. Nordberg & Max Costa eds., 2022).

On October 20, 2023, the same licensing requirements were announced to be extended to graphite, with an effective date of December 1, 2023.[54] On August 15, 2024, additional restrictions were declared on antimony and superhard materials, effective from September 15, 2024.[55] Both graphite and antimony compounds are potential semiconductor materials for a variety of optoelectronic applications.[56] By this time, all materials covered by the previous bans will be unable to pass through customs without proof of approval from the Ministry of Commerce, which will be required during customs clearance procedures.

On December 3, 2024, merely a day after the United States announced its updated export controls on China, extending these from defense-based to military-civil fusion applications, China significantly enhanced its prior bans with immediate effect.[57] Under the new measures, all dual-use items were prohibited from export to U.S. military uses and military users. Furthermore, gallium, germanium, antimony and superhard materials intended for dual-use purposes were "in principle" prohibited for export to the United

---

50. Madhav Patel & Athanasios K. Karamalidis, *Germanium: A Review of Its US Demand, Uses, Resources, Chemistry, and Separation Technologies*, SEPARATION AND PURIFICATION TECH. 1, 1 (2021).

51. Notice on Gallium and Germanium July 2023, *supra* note 7.

52. *Id.*

53. *Id.*

54. Notice on Graphite October 2023, *supra* note 7.

55. Notice on Antimony August 2024, *supra* note 7.

56. Ashok Srivastava & Naheem Olakunle Adesina, Graphene—Technology and Integration with Semiconductor Electronics, 21 THEORETICAL & COMPUTATIONAL CHEMISTRY 1, 21 (2022); Hang Bai, Yufang Li, Honglie Shen, Long Wang, Hechao Li, Zhihong Xie, Andi Chen, Zheng Shi, & Wei Wang, Preparation of Antimony Selenide Thin Films by Electrochemical Deposition and Application in Optoelectronic Devices, in MATERIALS SCI. IN SEMICONDUCTOR PROCESSING 1, 1 (2024).

57. Notice on Dual-Use Substance December 2024, *supra* note 7.

States, while the export of dual-use graphite became subject to more rigorous examination.[58]

The authority of the Ministry of Commerce to enforce these measures is granted by the Export Control Law of the People's Republic of China, promulgated in October 2020 and effective from December 2020.[59] The law states its objective as "safeguarding national security and interest"[60] and signals the retaliatory nature of export control by explicitly providing that China may take "reciprocal measures" against any country or region that "abuses export control measures to endanger the national security and national interests of [China]."[61] Under the Export Control Law, the export of dual-use items is already subject to the licensing requirements. Nevertheless, the December 2024 measure prohibited all dual-use items, even those not previously identified as essential to national security interests, from being exported to U.S. military uses or users. Additionally, the export of materials that were already regulated is now either subject to more stringent examination or entirely banned. This updated measure clearly signals a strong retaliatory intent. Apart from announcing the measure merely a day after the United States tightened its control with immediate effect, the Ministry of Commerce explicitly stated that this export control, unlike previous ones, specifically targeted the United States.

Export controls have also been leveraged during the latest round of tariff war between the United States and China. On April 4, 2025, China further announced control measures on seven heavy REEs, including samarium, gadolinium, terbium, dysprosium, lutetium, scandium and yttrium.[62] A stable supply of these elements is crucial for the production of products such as electric vehicles, robots, wind turbines, jet engines and nuclear submarines.[63] Under the new regulation, the export of all these elements requires licenses, and the exporter is obliged to determine whether an exported item falls under the category of regulated materials.[64] Reportedly, China approved export permits following the trade talks between high-level officials held in London in June 2025.[65]

---

58. *Id.*

59. Chukou Guanzhi Fa (出口管制法) [Export Control Law], (promulgated by the Standing Comm. Nat'l People's Cong., Oct. 17, 2020, effective Dec. 1, 2020) 2020 Standing Comm. Nat'l People's Cong. Gaz. 778 (China).

60. *Id.* at art. 1.

61. *Id.* at art. 48.

62. Notice on Heavy Rare Earths April 2025, *supra* note 7.

63. Josh Funk & Didi Tang, *China Grants Rare Earth Export Permits after US Trade Talks, Offers Relief but Uncertainty Persists*, AP NEWS (June 12, 2025), https://apnews.com/article/rare-earths-china-united-states-trade-supply-chain-de92222cda02dc85064c697911c6dea7.

64. *Id.*

In summary, the past several years have witnessed an escalation of export controls between the United States and China. As Figure 1 below shows, while the U.S. export controls initiated from the semiconductor field and later spanned into AI models and investment, China's export controls were mainly focused on minerals and rare earths, which were both raw materials of semiconductor chips.

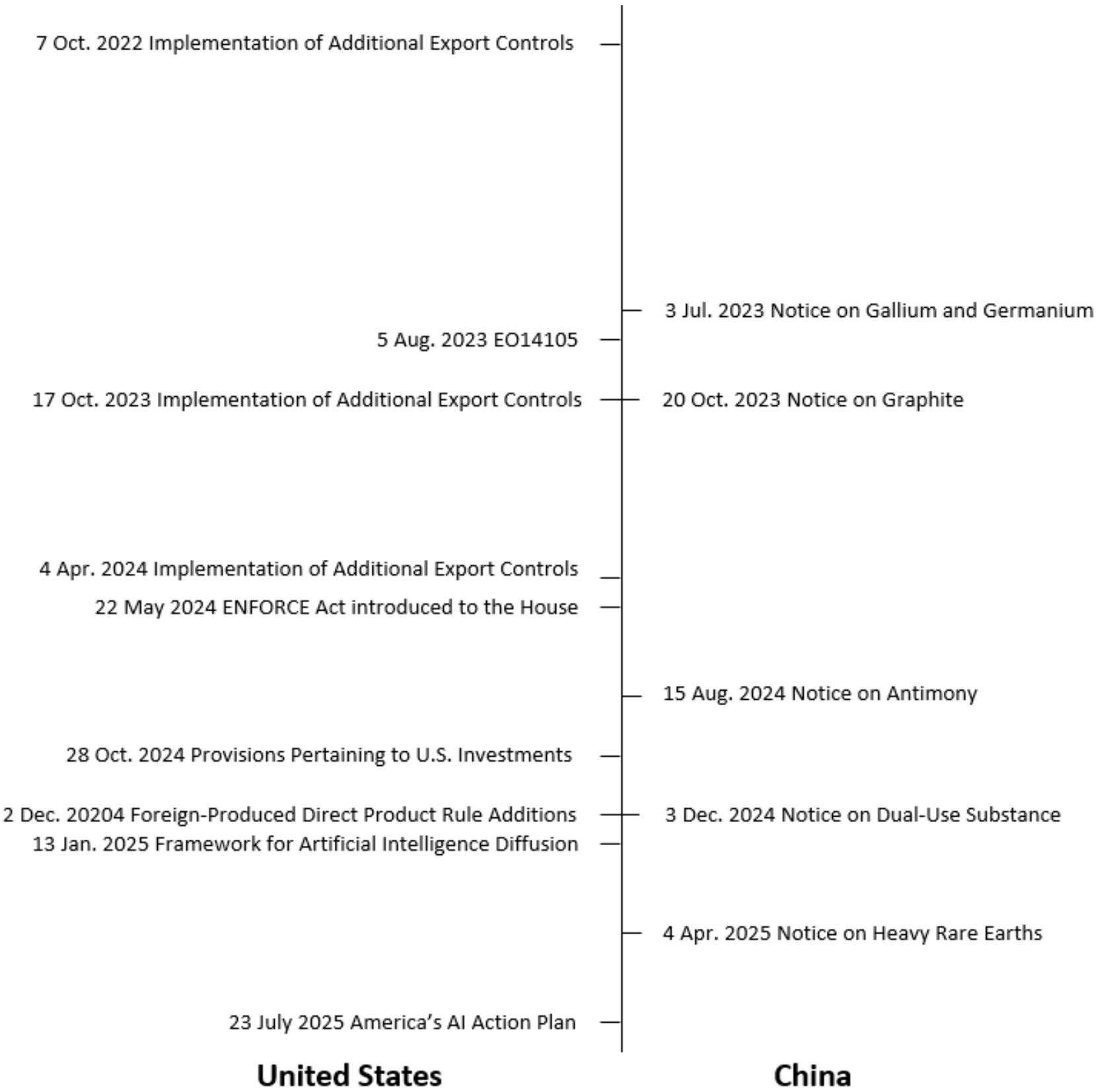

Figure 1 – Export control measures adopted by the United States and China

65. *Id.*

## III. CONSEQUENCE OF THE RESTRICTIONS

While both countries adopted the export restriction measures with an intention to curb the AI development of their adversary, these measures may turn out less effective than expected. This result is due to both countries' ability to diversify supply chain in the short run and finance independent R&D in the long run.

### *A. Alternative Sources*

1. The United States

U.S. officials and industrial stakeholders have recognized that China's control of critical minerals would yield short-term repercussions.[66] China accounts for over 95% of the global supply of gallium,[67] over 80% of germanium,[68] over 65% of graphite,[69] over 60% of antimony,[70] over 70% of samarium,[71] gadolinium,[72] terbium, and dysprosium,[73] over 95% of lutetium,[74] and over 70% of scandium[75] and yttrium.[76] Beyond domestic mining, China also exerts control over the global supply of these elements by monopolizing the refining and processing of heavy rare earths[77] and

---

66. Idrees Ali et al., *Pentagon Has Strategic Germanium Stockpile but No Gallium Reserves*, REUTERS (July 6, 2023), https://www.reuters.com/markets/commodities/pentagon-has-strategic-germanium-stockpile-no-gallium-reserves-2023-07-06/.

67. Zhongkui Han, Qiance Liu, Xin Ouyang, Huiling Song, Tianming Gao, Yanfei Liu, Bojie Wen & Tao Dai, *Tracking Two Decades of Global Gallium Stocks and Flows: A Dynamic Material Flow Analysis*, 202: 107391 RESOURCES, CONSERVATION & RECYCLING 1, 6 (2024).

68. Patel & Karamalidis, *supra* note 50, at 2.

69. Francis Isidore Barre, Romain Guillaume Billy, Fernando Aguilar Lopez & Daniel Beat Müller, *Limits to Graphite Supply in a Transition to a Post-Fossil Society*, 208: 107709 RESOURCES, CONSERVATION & RECYCLING 1, 1 (2024).

70. Susan Van den Brink, René Kleijn, Benjamin Sprecher, Nabeel Mancheri & Arnold Tukker, *Resilience in the Antimony Supply Chain*, 186: 106586 RESOURCES, CONSERVATION & RECYCLING 1, 1 (2022).

71. Critical Materials Monitor, COLUM. UNIV. CTR. ON GLOB. ENERGY POL'Y (2024), https://criticalmaterials.energypolicy.columbia.edu/minerals/Sm/.

72. Critical Materials Monitor, COLUM. UNIV. CTR. ON GLOB. ENERGY POL'Y (2024), https://criticalmaterials.energypolicy.columbia.edu/minerals/Gd/.

73. Critical Materials Monitor, COLUM. UNIV. CTR. ON GLOB. ENERGY POL'Y (2024), https://criticalmaterials.energypolicy.columbia.edu/minerals/Dy/.

74. *Lutetium Oxide Market Size, Share, Growth and Industry Analysis*, GLOB. GROWTH INSIGHTS (July 21, 2025), https://www.globalgrowthinsights.com/market-reports/lutetium-oxide-market-108855.

75. Critical Materials Monitor, COLUM. UNIV. CTR. ON GLOB. ENERGY POL'Y (2024), https://criticalmaterials.energypolicy.columbia.edu/minerals/Sc/.

76. Critical Materials Monitor, COLUM. UNIV. CTR. ON GLOB. ENERGY POL'Y (2024), https://criticalmaterials.energypolicy.columbia.edu/minerals/Y/.

77. *China Has a Weapon That Could Hurt America: Rare-Earth Exports*, THE ECONOMIST (Apr. 10, 2025), https://www.economist.com/finance-and-economics/2025/04/10/china-has-a-weapon-that-could-hurt-america-rare-earth-exports.

acquiring stakes in foreign mines.[78] In recent years, the United States has heavily depended on China as the primary exporter of these substances.[79] The restrictions imposed on those elements, contingent upon the stockpile and reserves available, will directly affect the United States' capacity and cost of producing advanced AI-embedded systems.

Despite China's unequivocal dominance in the supply and processing of rare earths, other resource-rich countries perceive the rift between the United States and China in the mineral market as an opportunity. For example, the Australian government has resolved to expand its critical minerals and rare earths industry to diversify the global supply chain and "support the security needs of Australia and *its partners*."[80] In addition to funding from their own government, Australian miners have also attracted interest from the U.S. Export-Import Bank, a federal agency, offering an investment of up to USD 600 million.[81] Concurrently, the Australian government is investing in rare earth refining and processing technologies as part of its "Future Made in Australia" initiative.[82] Immediately following China's announcement of controls on seven REEs in April 2025, Australia's re-elected Labour Government unveiled a plan to establish a Critical Minerals Strategic Reserve, with an initial investment of AUD 1.2 billion (approximately USD 785 million).[83] This initiative followed negotiations between Australia and the Trump Administration concerning a potential agreement on these minerals for a tariff exemption.[84] Currently, Australia supplies around six percent of the world's REEs.[85]

---

78. *See, e.g.*, Tim Greitemeier, Achim Kampker, Jens Tübke & Simon Lux, *China's Hold on the Lithium-ion Battery Supply Chain: Prospects for Competitive Growth and Sovereign Control*, 32: 100173 J. POWER SOURCES ADVANCES 1, 3 (2025) (attributing China's control over several critical elements to Chinese companies' acquisition of foreign mining rights).

79. U.S. DEP'T OF THE INTERIOR, U.S. GEOLOGICAL SURV., MINERAL COMMODITY SUMMARIES 2024 (2024).

80. Madeleine King MP, *Supporting Rare Earths Processing for a Future Made in Australia*, AUSTL. GOV'T MIN. OF INDUS., SCI. & RESOURCES (Oct. 15, 2024), http://www.minister.industry.gov.au/ministers/king/media-releases/supporting-rare-earths-processing-future-made-australia.

81. *Id.*

82. Anthony Albanese MP & Jim Chalmers MP, *Investing in a Future Made in Australia*, PRIME MIN. OF AUSTL. (May 14, 2024), https://www.pm.gov.au/media/investing-future-made-australia.

83. Anthony Albanese PM, *Albanese Government to Establish Critical Minerals Strategic Reserve*, ANTHONY ALBANESE ELECTORATE OFF. (Apr. 24, 2025), https://anthonyalbanese.com.au/media-centre/albanese-government-to-establish-critical-minerals-strategic-reserve.

Canadian firms are likewise seizing the opportunity to forge more agreements with their southern neighbor. The Saskatchewan Research Council, established by the provincial government, is in the process of constructing a Rare Earth Processing Facility within its territory, which successfully produced rare earth metals at a commercial scale in 2024, marking the first jurisdiction in North America to achieve such a feat.[86] The Facility has received significant provincial and national funding, aiming to "build a strong, sustainable rare earth supply chain for Canada and *[its] allies*."[87] In addition to primary processing, Canada possesses recycling capabilities for REE. In April 2025, Canadian REE recycling firm Cyclic Materials secured a Series B investment from Amazon to support "the expansion of Cyclic Materials' innovative recycling technologies, [thereby] strengthening the circular supply chain for critical materials used in data centers and other industries."[88] However, in light of the escalating tension and ongoing tariffs between Canada and the United States, commentators have noted the potential for a future export ban on these elements from Canada as well, which could ultimately benefit China in its bilateral negotiations with the United States.[89]

In the meantime, the United States is engaged in negotiations with its European and African partners. In April 2025, the United States entered into an agreement with Ukraine for joint investment in the exploitation of Ukrainian natural resources, particularly critical REEs.[90] This military-economic agreement, awaiting approval from Ukrainian lawmakers, grants the United States preferential access to new mineral extraction opportunities in Ukraine in exchange for American military support in Ukraine's war with Russia.[91] Following the initial announcement of this potential deal in February 2025, Russia reportedly extended an offer to the United States for joint exploration of REEs and other metals within its territory.[92] Notably, Ukraine claims to possess around five percent of the world's "critical raw materials,"[93] while Russia possesses the fifth-largest reserves of rare earth metal globally.[94]

---

84. Michelle Grattan, *Albanese Government Announces $1.2 billion Plan to Purchase Critical Minerals*, THE CONVERSATION (Apr. 23, 2025), https://theconversation.com/albanese-government-announces-1-2-billion-plan-to-purchase-critical-minerals-254994.

85. Ian Johnston, Alice Hancock, Harry Dempsey & Visual Storytelling Team, *Can Europe Go Green without China's Rare Earths?*, FIN. TIMES (Sept. 20, 2023), https://ig.ft.com/rare-earths/.

86. *SRC Rare Earth Processing Facility First to Produce Rare Earth Metals in North America*, SASKATCHEWAN RSCH COUNCIL (Sept. 16, 2024), https://www.src.sk.ca/news/src-rare-earth-processing-facility-first-produce-rare-earth-metals-north-america.

87. *Id.*

88. *Amazon Invests in Cyclic Materials' Oversubscribed Series B Equity Round*, CYCLIC MATERIALS, https://www.cyclicmaterials.earth/resources/amazon-invests-in-cyclic-materials-oversubscribed-series-b-equity-round, (last visited Aug. 7, 2025).

89. Daniel, *Canada Responds with Tariffs of its Own – What are the Ramifications for Critical Minerals and Rare Earth Elements?*, RARE EARTH EXCH. (Feb. 3, 2025), https://rareearthexchanges.com/news/canada-u-s-trade-conflict/.

90. The Full Text of the Ukraine-US Minerals Agreement, EUR. PRAVDA (Feb. 26, 2025), https://www.eurointegration.com.ua/eng/articles/2025/02/26/7205922/.

In Africa, the U.S. International Development Finance Corporation (DFC) is investing in South Africa's Phalaborwa Rare Earths Project, with a proposed equity amount of USD 50 million, to bolster South Africa's production and export of rare earth oxides.[95] The Democratic Republic of Congo (DRC) has also approached the United States regarding a potential mineral-for-peace agreement akin to the one struck with Ukraine.[96] Nevertheless, both projects face significant uncertainty due to the recent strain on the relations between South Africa and the United States, as well as the DRC's historical economic reliance on China.[97]

In Asia, Vietnam holds approximately four percent of the world's rare earth reserves.[98] In September 2023, Vietnam and the United States announced an elevation in their relationship to a Comprehensive Strategic Partnership, underscored by agreements on semiconductors and minerals. The two nations jointly proclaimed:

> [r]ecognizing Vietnam's enormous potential to become a key country in the semiconductor industry, the two leaders supported the rapid development of Vietnam's semiconductor ecosystem and pledged to

---

91. *Id.*
92. Ola Noureldin, *Putin Proposes Joint Rare Earth Metals Exploration With US, Offers Aluminum Supply*, FORBES (Feb. 25, 2025), https://www.forbesmiddleeast.com/industry/energy/putin-offers-us-joint-rare-earth-metals-exploration-and-aluminum-supply.
93. Ian Aikman & João da Silva, *What We Know about US-Ukraine Minerals Deal*, BBC (April 30, 2025), https://www.bbc.com/news/articles/cn527pz54neo.
94. Noureldin, *supra* note 92.
95. U.S. INT'L DEV. FIN. CORP., TECHMET PHALABORWA RARE EARTHS (2023).
96. Damian Zane, *Is Trump Mulling a Minerals Deal with Conflict-Hit DR Congo?*, BBC (Mar. 11, 2025), https://www.bbc.com/news/articles/cjryjlndddpo.
97. *Id.*; Yeganeh Torbati, *Trump's Feud with South Africa Could Hinder Hunt for Critical Minerals*, THE WASH. POST (Apr. 30, 2025), https://www.washingtonpost.com/business/2025/04/30/trump-china-minerals-south-africa-investment/.
98. Caroline Peachey, *Exploring US Efforts to Find a Secure Supply of Rare Earth Elements*, MINING TECH. (May 9, 2025), https://www.mining-technology.com/analysis/exploring-us-efforts-to-find-a-secure-supply-of-rare-earth-elements/?cf-view&cf-closed.

> actively cooperate to enhance Vietnam's position in the global semiconductor supply chain.[99]

As part of the Partnership, the United States agreed to provide an initial seed grant of USD 2 million to support the development of Vietnam's semiconductor workforce.[100] An additional USD 2 million was allocated to STEM-related education and training.[101] According to the U.S. government, the Partnership has fostered enhanced cooperation between the two nations concerning economic security, semiconductor development, and critical minerals.[102] In July 2024, a joint initiative was established upon among American, South Korean and Vietnamese stakeholders to collaboratively develop rare earth mines in Vietnam.[103]

Concurrently, various countries are exploring innovative methods to extract rare earths to mitigate their reliance on China. In July 2025, Japan announced its collaboration with the United States, India, and Australia to embark on extracting rare earth minerals from the ocean floor, starting from January 2026.[104] Consequently, deep-sea mining will become a key area for rare-earth R&D alongside direct mining and recycling.

Despite the availability of various alternative sources, reliance on imports from foreign countries fails to offer a lasting solution to the scarcity of critical materials and REEs in the United States. Predominantly, China accounts for nearly 70 percent of global REE production, a figure that eclipses the combined output of all other countries. Beyond its natural abundance, China commands 90 percent of the processing capabilities. Although countries such as Australia have initiated their mining endeavors, most of these projects remain in developing stages and are not expected to yield output for several years. Moreover, for the U.S. semiconductor industry, procuring raw materials from these other countries would not only escalate production costs but also introduce significant uncertainty, as any foreign supply chain is susceptible to burgeoning geopolitical risks. Therefore, as will be illustrated below,[105] in addition to finding alternative foreign suppliers, there has been a concerted effort within the United States to establish a domestic supply chain through investing in domestic mining and refining technologies.

---

99. *Việt Nam, Joint Statement on Upgrading Vietnam - US Relations to Comprehensive Strategic Partnership*, VIETNAM.VN (Sept. 12, 2023), https://www.vietnam.vn/en/toan-van-tuyen-bo-chung-ve-nang-cap-quan-he-viet-nam-hoa-ky-len-doi-tac-chien-luoc-toan-dien.

100. *Id.*

101. *Id.*

102. Press Release, U.S. Dep't of State, U.S. Relations with Vietnam (Jan. 10, 2025), https://2021-2025.state.gov/bureau-of-east-asian-and-pacific-affairs/releases/2025/01/u-s-relations-with-vietnam/.

103. Thanh Van, *South Korea's Trident Forms Joint Venture with US' Zoetic to Develop Rare Earth Mines in Vietnam*, VIETNAM INV. REV. (July 29, 2024), https://vir.com.vn/south-koreas-trident-forms-joint-venture-with-us-zoetic-to-develop-rare-earth-mines-in-vietnam-113068.html.

104. Yuka Obayashi, *Japan to Begin Test Mining Rare-Earth Mud from Seabed in Early 2026*, REUTERS (July 4, 2025), https://www.reuters.com/markets/asia/japan-begin-test-mining-rare-earth-mud-seabed-early-2026-2025-07-04/.

2. China

China's quest to identify alternative sources of semiconductor technology is decidedly more arduous than that of the United States. First, not only U.S. firms but also the allies of the United States, such as ASML from the Netherlands and Tokyo Electron from Japan, have joined or been forced to join the initiative to prohibit the export of certain semiconductor manufacturing equipment to China.[106] Furthermore, even if China succeeds in securing supplies from other countries, Nvidia's GPU designs are widely acknowledged as the industry standard.[107] By restricting Nvidia from selling its most advanced models to Chinese customers, the U.S. government can effectively control China's access to cutting-edge semiconductor chips and chipmaking technologies.

Despite these challenges, China has somewhat mitigated the impact of U.S. sanctions while buying time to expedite the development of its domestic chipmaking industry. According to the statistics released by Chinese Customs, imports of semiconductor machinery into China surged to USD 44.4 billion in 2024, marking a year-on-year increase of 18.1%.[108] The Netherlands and Japan remained the primary exporters, with a market share of 29.2% and 22%, respectively.[109] Chinese Customs highlighted that the main driver for this growth in imports was local semiconductor firms increasing their stockpiles to manage "external risks."[110]

---

105. *See* infra Part II.2.A.

106. Statement Regarding Partial Revocation Export License, ASML (Jan. 1, 2024), https://www.asml.com/en/news/press-releases/2023/statement-regarding-partial-revocation-export-license; Press Conference by Minister Nishimura (Excerpt), MIN. OF ECON., TRADE & INDUS. OF JAPAN (Mar. 31, 2023), https://www.meti.go.jp/english/speeches/press_conferences/2023/0331001.html.

107. Dylan Butts, *China's Racing to Build Its AI Chip Ecosystem as U.S. Curbs Bite. Here's How Its Supply Chain Stacks Up*, CNBC (June 11, 2025), https://www.cnbc.com/2025/06/12/chinas-racing-to-beat-us-chip-curbs-how-its-supply-chain-stacks-up.html?msockid=14788734181869d83c89920119c1682e.

108. Yi He, *2024 Nian Woguo Bandaoti Zhizao Shebei Jinkou E Lianxu Di Er Nian Chuang Xingao* (2024年我国半导体制造设备进口额连续第二年创新高) [*China's Semiconductor Manufacturing Equipment Imports Hit a Record High for the Second Consecutive Year in 2024*] CHINA CHAMBER OF IMP. & EXP. OF MACH. & ELEC. PROD. (Feb. 7, 2025), https://www.cccme.org.cn/news/details.aspx?id=F650BFD8BE469D6B672EC7C07302E618&classid=7685F2FA1E54A7F5&xgid=F868932F64EB7AAF.

109. *Id.*

As U.S. export controls have shown a widening scope in recent years, Chinese companies have strategized by accumulating stocks of lower-end or older-generation models before updates to restrictions.[111] For example, the success of DeepSeek has been partly attributed to its acquisition of a substantial quantity of Nvidia A100 GUPs prior to the enforcement of export restrictions targeting them.[112] DeepSeek's research team's recent publication also revealed that DeepSeek-V3 was trained on 2,048 Nvidia H800 GPUs, which remained legal for export until 2023.[113] Their success prompted further U.S. government bans on these lower-end chips tailored for the Chinese market in April 2025.[114]

Another controversy surrounding DeepSeek involves allegations that it bypassed export controls by acquiring Nvidia chips through third parties. It is particularly speculated that DeepSeek channeled restricted Nvidia chips via intermediaries in Singapore, Malaysia and the United Arab Emirates (UAE), a practice now so widespread that it is often referred to as "Singapore washing."[115] In response, the Singaporean Ministry of Trade and Industry, wary of being placed under stricter scrutiny in the U.S. export control system, issued a statement clarifying that many purchases made by

---

110. *Id.*

111. *Chinese Imports of Chip Gear Hit Record US$26 Billion This Year*, BLOOMBERG (Aug. 22, 2024), https://www.bloomberg.com/news/articles/2024-08-22/chinese-imports-of-chip-gear-hit-record-26-billion-this-year; Iris Deng, *Tech War: China's Chip Imports Surge as Firms Stockpile Ahead of Fresh US Restrictions*, S. CHINA MORNING POST (Aug. 7, 2024), https://www.scmp.com/tech/tech-war/article/3273606/tech-war-chinas-chip-imports-surge-firms-stockpile-ahead-fresh-us-restrictions?module=perpetual_scroll_0&pgtype=article.

112. Ritwik Gupta, Leah Walker & Andrew W. Reddie, *Whack-A-Chip: The Futility of Hardware-Centric Export Controls*, AI FRONTIERS INITIATIVE 1, 4 (Nov. 21, 2024), https://arxiv.org/pdf/2411.14425; Malik Sallam et al., *DeepSeek: Is It the End of Generative AI Monopoly or the Mark of the Impending Doomsday?*, MESOPOTAMIAN J. BIG DATA 26, 27 (2025); Khushboo Razdan & Mark Magnier, *US Tech Stocks Plunge on China AI's Unexpectedly Strong Showing*, S. CHINA MORNING POST (Jan. 28, 2025), https://www.scmp.com/news/china/article/3296510/us-tech-stocks-plunge-china-ais-unexpectedly-strong-showing.

113. Chenggang Zhao et al., *Insights into DeepSeek-V3: Scaling Challenges and Reflections on Hardware for AI Architectures*, (The 52nd Ann. Int'l Symp. on Comput. Architecture, Tokyo, Japan, June 2025), https://doi.org/10.48550/arXiv.2505.09343.

114. Press Release, H. R. Select Comm. on the CCP, Moolenaar, Krishnamoorthi Call for Tightening Export Controls on Chips Critical to China's AI Platform DeepSeek and Other Measures to Address its Risks to Americans Data and Security (Jan. 30, 2025), https://chinaselectcommittee.house.gov/media/press-releases/moolenaar-krishnamoorthi-call-for-tightening-export-controls-on-chips-critical-to-china-s-ai-platform-deepseek-and-other-measures-to-address-its-risks-to-americans-data-and-security. [hereinafter Krishnamoorthi Call for Tightening Export Controls].

115. James Kynge, Jude Webber & Christine Murray, *"Singapore Washing": China's New Back Doors into Western Markets*, FIN. REV. (Sept. 9, 2024), https://www.afr.com/world/asia/singapore-washing-china-s-new-back-doors-into-western-markets-20240909-p5k8yj.

Singaporean entities were destined for Western countries, and affirmed its ongoing close cooperation with U.S. authorities.[116] Additionally, three individuals, including one Chinese national, were arrested in Singapore for mislabeling the end user of restricted Nvidia chips.[117] Similarly, the Malay Minister of Investment, Trade and Industry also declared that the Malay government is "enhancing international cooperation to monitor the flow of sensitive technologies, such as Nvidia's H100 and A100 chips."[118]

However, even with the backing from the Singaporean and Malay governments, the United States cannot ensure semiconductor chips dispatched to these countries will not eventually reach China or Chinese entities, as the actions of private stakeholders are difficult to control. The U.S. government has been attempting to address this loophole from two angles. From the seller's side, the House Select Committee on the CCP has launched an investigation against Nvidia, requesting the company to disclose all communications between Nvidia and DeepSeek, as well as information and contracts with all customers located or listed an end use in China (including Hong Kong and Macau) and ASEAN member states (i.e., Brunei, Cambodia, Indonesia, Laos, Malaysia, Myanmar, the Philippines, Singapore, Thailand and Vietnam) since 2020.[119] In a letter to Nvidia Chief Executive Officer Jensen Huang, the Committee highlighted that:

> Despite multiple rounds of U.S. export restrictions on AI chips, DeepSeek's ability to develop advanced AI models suggests that loopholes or indirect supply channels may still exist. This raises significant questions about the effectiveness of current U.S. export control laws in preventing the transfer of sensitive AI technologies to adversarial regimes. If not addressed, such gaps could undermine the United

---

116. Press Statement on Whether DeepSeek Gained Access to US Export-Controlled Chips through Intermediaries in Singapore, SING. MIN. OF TRADE & INDUS. (Feb. 2025), https://www.mti.gov.sg/Newsroom/Press-Releases/2025/02/Press-Statement-on-whether-DeepSeek-gained-access-to-US-export-controlled-chips.

117. *3 Men Charged with Fraud, Cases Linked to Alleged Movement of Nvidia Chips*, CNA (Feb. 27, 2025), https://www.channelnewsasia.com/singapore/3-men-charged-fraud-nvidia-chips-singapore-china-deepseek-4964721.

118. *Malaysia Takes Nvidia AI Chip Smuggling Claims to China Seriously, Reaffirms Export Control Compliance, Minister Says*, MALAY MAIL (Feb. 7, 2025), https://www.malaymail.com/news/malaysia/2025/02/07/malaysia-takes-nvidia-ai-chip-smuggling-claims-to-china-seriously-reaffirms-export-control-compliance-minister-says/165802.

119. H. R. Select Comm. on the CCP, Nvidia Letter (Mr. Jensen Huang, CEO) (Apr. 16, 2025).

> States' technological leadership, with far-reaching *economic and national security implications*.[120]

From the buyer's perspective, concern over indirect supply routes is one reason behind the BIS' periodic update of its control measures, particularly regarding the country lists. In another letter to the U.S. National Security Advisor addressing this matter, the Committee stated "[c]ountries like Singapore should be subject to strict licensing requirements absent a willingness to crack down on PRC transshipment through their territory."[121]

Nevertheless, both approaches have their drawbacks: while banning Nvidia from sales to China has significant repercussions on U.S. technology stocks,[122] periodically expanding the control list to include more countries will not fully resolve the issue until every nation is categorized under the most restrictive tier. Therefore, as acknowledged by the Committee itself, the persistence of these possible circumvention routes casts doubts on the effectiveness of the export controls, particularly given that these controls come at the cost of impacting the United States' own economy.

### *B. Independent R&D*

#### 1. The United States

As mentioned above, procuring raw materials from other countries may not be able to provide a long-term solution to the U.S. semiconductor industry. Recognizing the peril of dependency on foreign minerals, President Trump proclaimed a national energy emergence via an executive order on January 20, 2025.[123] In response to the declared emergency, he signed another executive order on March 20, 2025 amid the ongoing tariff conflict, which introduced immediate measures to bolster American mineral production.[124] These initiatives include prioritizing mineral extraction and mining as a principal land use in designated areas, identifying mineral production as a critical area for industrial capability development, and providing government loans, capital assistance, technical support, and working capital to sponsors of domestic mineral production projects.[125] Additionally, the measures promote the amalgamation of private capital with domestic mineral ventures and the establishment of a dedicated fund for domestic investments in minerals, managed by the DFC.[126]

---

120. *Id.*
121. Krishnamoorthi Call for Tightening Export Controls, *supra* note 114.
122. *See, e.g.*, George Steer & Will Schmitt, *Tech Stocks Sink after Nvidia Reveals Hit from US Curbs on Sales to China*, FIN. TIMES (Apr. 16, 2025), https://www.ft.com/content/0e18bb8e-30cd-448d-99cb-98b266753391.
123. Exec. Order 14156, 90 Fed. Reg. 8433 (Jan. 20, 2025).
124. Exec. Order 14241, 90 Fed. Reg. 13673 (Mar. 20, 2025).

According to both executive orders, the "critical minerals" encompasses any mineral, element, substance, or material identified as critical by the Secretary of the Interior, acting through the Director of the United States Geological Survey (USGS). This designation is based on the criteria that they are deemed "essential to the *economic and national security* of the United States," possessing a vulnerable supply chain, and crucial to manufacturing a product.[127] Following these criteria, the 2022 list published by the USGS designated 50 substances as "critical," with China being the leading producer of 29 of these substances.[128] Notably, all 11 substances recently subjected to China's export controls were included in this list.[129] As foreseen by industry experts, the imposition of export controls by China has spurred U.S. legislators to intensify investments in critical minerals, aiming for self-sufficiency.[130]

Currently, the sole operational REE mine in the United States is Mountain Pass in California. The discovery of rare earths at Mountain Pass occurred in 1949, and subsequent to this, a company secured a mining claim in the area, initiating modest production in 1952.[131] However, mining operations ceased in 2002, and, by 2015, the mine was idled, partly due to fierce competition from China and insufficient investment domestically.[132] In 2017, a new company, MP Materials, was established to assume control of the mine, and it subsequently went public on the New York Stock Exchange in 2020.[133] The company's mission is articulated as "restoring the full U.S. rare earth supply chain to *support the energy transition and strengthen national security*."[134] By 2022, it contributed approximately 15 percent to global rare earth oxides production.[135] However, the mission continues to face significant challenges. First, upon resumption of operations, raw rare earth concentrates extracted at Mountain Pass were transported directly to China for processing, a practice which persisted until China tightened its export controls on REEs in 2025.[136] Second, the involvement of a Chinese state-owned enterprise (SOE), Shenghe Resources, holding around 8 percent of MP Materials' shares, introduces geopolitical complexities into the supply chain.[137] Third, California's stringent environmental protection laws render domestic mining and processing both challenging and costly.[138]

---

125. *Id.*
126. *Id.*
127. Energy Act of 2020, H. R. 4447, 116th Cong. § 7002(a)(2) (2020).
128. CONG. RSCH. SERV., CRITICAL MINERAL RESOURCES: THE U.S. GEOLOGICAL SURVEY (USGS) ROLE IN RESEARCH AND ANALYSIS (Feb. 21, 2025).
129. U.S. DEP'T OF THE INTERIOR, GEOLOGICAL SURVEY 2022 FINAL LIST OF CRITICAL MINERALS (2022) (The critical materials designated by the USGS include antimony, dysprosium, gadolinium, gallium, germanium, graphite, lutetium, samarium, scandium, terbium and yttrium).
130. Ali et al., *supra* note 66.
131. MP Materials Corp., History, MP MATERIALS, https://mpmaterials.com/history, (last visited June 24, 2025).
132. *Id.*
133. *Id.*
134. MP Materials Corp., *Founder-Led and Owner-Operated*, MP MATERIALS, http://mpmaterials.com/about/, (last visited June 24, 2025).

In response to these concerns, the United States is implementing measures that support the research, development and production at Mountain Pass, while simultaneously working to diversify MP Materials' supply chain, capital source, and revenue streams to reduce its reliance on China. In December 2021, General Motors entered into a long-term agreement with MP Materials, positioning itself as a primary consumer of the miner's output.[139] In February 2022, the Department of Defense (DoD) awarded MP Materials a USD 35 million contract to construct a more sustainable facility for processing heavy REEs at Mountain Pass, aimed at supporting both defense and commercial applications within the United States.[140] In April 2024, the company further received a USD 58.5 million award in tax credits from a program administered by the Department of Energy (DoE).[141] In July 2025, Apple announced a multi-year commitment of USD 500 million to buy American-made rare earth magnets from MP Materials.[142] Furthermore, the U.S. government is actively promoting the development of a second rare earth mine, the Colosseum mine, also in California, in collaboration with an Australian mining company.[143]

---

135. *The Mountain Pass Mine in California May Be the U.S. Rare Earths Game Changer*, CAL. CURATED (Jan. 29, 2025), https://californiacurated.com/2025/01/29/the-mountain-pass-mine-in-california-may-be-the-u-s-rare-earths-game-changer/.

136. *Id.*

137. *Id.*

138. Ruth Jebe, Don Mayer & Yong-Shik Lee, *China's Export Restrictions of Raw Materials and Rare Earths: A New Balance Between Free Trade and Environmental Protection?*, 44 GEO. WASH. INT'L L. REV. 101 (2012); Julia Ya Qin, *The China-Raw Materials Case and Its Impact (or Lack Thereof) on U.S. Downstream Industries*, 106 PROC. AM. SOC'Y INT'L L. 278, 280 (2012).

139. CAL. CURATED, *supra* note 135.

140. Press Release, U.S. Dep't of Defense, DoD Awards $35 Million to MP Materials to Build U.S. Heavy Rare Earth Separation Capacity (Feb.22, 2022), https://www.war.gov/News/Releases/Release/Article/2941793/dod-awards-35-million-to-mp-materials-to-build-us-heavy-rare-earth-separation-c/.

141. MP Materials Corp., *MP Materials Awarded $58.5 Million to Advance Earth Magnet Manufacturing*, MP MATERIALS (Apr. 1, 2024), https://investors.mpmaterials.com/investor-news/news-details/2024/MP-Materials-Awarded-58.5-Million-to-Advance-U.S.-Rare-Earth-Magnet-Manufacturing/default.aspx.

142. *Apple Expands U.S. Supply Chain with $500 Million Commitment to American Rare Earth Magnets*, APPLE NEWSROOM (July 15, 2025), https://www.apple.com/newsroom/2025/07/apple-expands-us-supply-chain-with-500-million-usd-commitment/.

Beyond direct mining and refining, REEs can also be sourced from recycling industrial and urban wastes.[144] IonicRE, an Australian company with a UK subsidiary that has developed and patented technologies for recycling permanent magnet, is planning to establish multiple recycling facilities across the United States.[145] The company's recycling process is capable of producing heavy REEs currently under China's export restrictions, including dysprosium, terbium, samarium and scandium.[146] In alignment with the U.S. government's aim to eliminate reliance on Chinese minerals in defense applications by 2027, the company anticipates governmental support in its strategic expansion.[147] Similarly, American company Tusaar is set to construct a recycling plant in Denver in 2025, with funding secured from both the DoD and the DoE.[148] Apple's recent deal with MP Materials also includes the launch of a new recycling facility for processing recycled REEs.[149] Notably, there is considerable untapped potential in rare earth recycling, as currently only around one percent of all rare earths in e-waste are recycled, largely due to inadequate capabilities and incentives.[150] With significant governmental investment and incentive mechanisms in place, recycling of rare earths, considered more environmentally friendly compared to direct mining, is expected to increase significantly in the United States. As recycling and primary mining of rare earths are complementary activities,[151] the United States' concurrent investment in both domains, if executed effectively, could mitigate the impact of China's export control on critical materials in the long run.

---

143. Staff Writer, *US Interior publicly backs rare earth mine next to Mountain Pass,* MINING.COM (June 10, 2025), https://www.mining.com/us-interior-publicly-backs-rare-earth-mine-next-to-mountain-pass/.

144. Yoshiko Fujita, *Recycling Rare Earths: Perspectives and Recent Advances*, 47 MRS BULL. 283, 283 (2022).

145. John Zadeh, *How Ionic Rare Earths Is Revolutionising Magnet Recycling in the USA*, DISCOVERY ALERT (June 23, 2025), https://discoveryalert.com.au/news/ionic-rare-earths-magnet-recycling-usa-2025/.

146. *Id.*

147. *Id.*

148. Tusaar Corp., *Tasaar Unveils Plans for $17 Million Rare Earth Magnet Recycling Plant in the United States*, TUSAAR CORP. (Jan. 17, 2024), https://tusaar.com/wp-content/uploads/2024/01/Tusaar-MRPR-01172024.pdf

149. APPLE NEWSROOM, *supra* note 142.

150. Maddie Stone, *One Problem for Renewables: Not Enough Rare Earths. One Solution: Recycling. But There's a Hitch*, BULL. OF THE ATOMIC SCI. (Apr. 22, 2024), https://thebulletin.org/2024/04/one-problem-for-renewables-not-enough-rare-earths-one-solution-recycling-but-theres-a-hitch/.

151. Binnemans et al., *Recycling of Rare Earths: A Critical Review*, 51 J. CLEANER PROD. 1, 19 (2013).

2. China

China's ambition to develop its domestic semiconductor industry predates the export controls imposed by the United States. The government has implemented specific measures, including various preferential taxing mechanisms and financing tools, to support this initiative.

In 2014, the State Council issued a Guidelines on Promoting National Integrated Circuit Development, under which the China Integrated Circuit Industry Investment Fund, colloquially known as "the Big Fund," was created.[152] This initiative was designed to bolster IC manufacturing within the country, with initial investments sourced from several large SOEs in telecommunications and finance sectors. The Big Fund was launched in two phases, the first in 2014 and the second in 2019. Beneficiaries of both phases include domestic chipmakers such as Semiconductor Manufacturing International Corporation (SMIC), an SOE, and Hua Hong, a key supplier to Huawei.

In July 2020, SMIC was publicly listed on the Shanghai Stock Exchange (SSE), with an initial investment of RMB 3.5 billion (approximately USD 488 million) from the Big Fund's Phase II. Between 2020 and 2021, SMIC engaged in multiple joint ventures with the Big Fund Phase II and municipal governments across China.[153] In June 2023, the Big Fund Phase II invested RMB 3 billion to Hua Hong during its initial public offering on the SSE.[154] Highlighting these advancements, in August 2023, Huawei introduced its Mate 60 Pro, which incorporates a seven-nanometer processor manufactured by SMIC, amidst the backdrop of the United States' export controls.

---

152. Guojia Jicheng Dianlu Chanye Fazhan Tuijin Gangyao (国家集成电路产业发展推进纲要) [National Guidelines on Promoting the Development of the Integrated Circuits Industry], (promulgated by the Min. of Comm. on June 26, 2014, effective June 26, 2014).

153. *Shasha Lai, Zhongxin Guoji Yu Guojia Dajijin Chengli Lingang Hezi Gongsi, Touzi Chao 80 Yi Meiyuan (*中芯国际与国家大基金成立临港合资公司，投资超80亿美元*)* [*SMIC and the Big Fund Entered into Joint Venture in Lingang, with Investment Exceeding USD8 Billion*], YICAI (Nov. 12, 2021), https://m.yicai.com/news/101227774.html; Yi Luo, *Zhongxin Guoji: Zhongxin Konggu, Guojia Jicheng Dianlu Jijin II He Yizhuang Guotou Gongtong Chengli Hezi Qiye* (中芯国际：中芯控股、国家集成电路基金II和亦庄国投共同成立合资企业) [*SMIC: SMIC Holding, National Integrated Circuit Fund II, and Etown Capital Established Joint Venture*], YICAI (Dec. 4, 2020), https://www.yicai.com/news/100865237.html; Gelonghui, *Zhongxin Guoji (688981.SH): Ni Xiang Guojia Jicheng Dianlu Jijin II Zhuanrang Zhongxin Shenzhen 22% De Guquan* (中芯国际(688981.SH)：拟向国家集成电路基金II转让中芯深圳22%的股权) [*SMIC(688981.SH): Intending to Transfer 22% of SMIC Shenzhen Shares to National Integrated Circuit Fund II*], WEBULL (Nov. 23, 2021), https://www.webull.com/news/47244459.

154. Zhi Wu, *30 Yi! Dajijin Erqi Chushou, Zhantou Huahong Bandaoti Hui A* (30亿！大基金二期出手，战投华虹半导体回A) [*3 Billion! Big Fund II Aids in Hua Hong's Return to A Share*], STCN (June 29, 2023), https://www.stcn.com/article/detail/904251.html.

In May 2024, the Big Fund launched its third phase, amassing registered capital of RMB 344 billion (approximately USD 47.09 billion), with investment from all six major commercial banks and the Chinese Ministry of Finance as investors.[155] While the ultimate recipients under Phase III remain undisclosed, it is noteworthy that SMIC, having benefited from both previous phases, is now advancing its technology towards producing more sophisticated five-nanometer chips.[156] Should this endeavor prove successful, China's domestic chipmaking capabilities would be positioned just one generation behind the forefront of three-nanometer technology.

In terms of taxation, China has been implementing preferential tax treatments for domestic IC enterprises since 2012 and constantly refining these measures up to 2019.[157] In July 2020, the State Council promulgated a policy document titled the Policies on Promoting the High-Quality Development of Integrated Circuit and

---

155. CCTV, *supra* note 9.

156. Qianer Liu, *China on Cusp of Next-Generation Chip Production Despite US Curbs*, FIN. TIMES (Feb. 6, 2024), https://www.ft.com/content/b5e0dba3-689f-4d0e-88f6-673ff4452977.

157. Guanyu Jingyibu Guli Ruanjian Chanye He Jicheng Dianlu Chanye Fazhan Qiye Suodeshui Zhengce De Tongzhi (关于进一步鼓励软件产业和集成电路产业发展企业所得税政策的通知) [Notice on Enterprise Income Tax Policies to Further Encourage the Development of Software and Integrated Circuits Industries], Notice of the Min. of Fin. (MoF) & State Taxation Admin. (STA) No. 27 [2012] (promulgated by the MoF & STA on Apr. 20, 2012, effective Jan. 1, 2011); Caizheng Bu, Guojia Shuiwu Zongju, Fazhan Gaige Wei, Gonghe He Xinxihua Bu Guanyu Jinyibu Guli Jicheng Dianlu Chanye Fazhan Qiye Suodeshui Zhengce De Tongzhi (财政部、国家税务总局、发展改革委、工业和信息化部关于进一步鼓励集成电路产业发展企业所得税政策的通知) [Notice on Enterprise Income Tax Policies to Further Encourage the Development of Integrated Circuits Industry], Notice of the MoF, STA, Nat'l Dev. & Reform Comm'n (NDRC) & the Min. of Indus. and Info. Tech. (MIIT) No. 6 [2015] (promulgated by the MoF, STA, NDRC & MIIT on May 4, 2016, effective Jan. 1, 2015); Caizheng Bu, Guojia Shuiwu Zongju, Fazhan Gaige Wei, Gonghe He Xinxihua Bu Guanyu Ruanjian He Jicheng Dianlu Chanye Qiye Suodeshui Youhui Zhengce Youguan Wenti De Tongzhi (财政部、国家税务总局、发展改革委、工业和信息化部关于软件和集成电路产业企业所得税优惠政策有关问题的通知), Notice on Some Issues Relating to Preferential Enterprise Income Tax Treatment for Software and Integrated Circuits Enterprises, Notice of the MoF, STA, NDRC & MIIT No. 49 [2016] (promulgated by the MoF, STA, NDRC & MIIT on May 4, 2016, effective Jan. 1, 2015); Caizheng Bu, Shuiwu Zongju, Guojia Fazhan Gaige Wei, Gongye He Xinxihua Bu Guanyu Jicheng Dianlu Shengchan Qiye Youguan Qiye Suodeshui Zhengce Wenti De Tongzhi (财政部、税务总局、国家发展改革委、工业和信息化部关于集成电路生产企业有关企业所得税政策问题的通知) [Notice on Some Issues Relating to Enterprise Income Tax Policies for Integrated Circuits Enterprises], Notice of the MoF, STA, NDRC & MIIT No. 27 [2018] (promulgated by the MoF, STA, NDRC & MIIT on Mar. 28, 2018, effective Jan. 1, 2018); Caizheng Bu, Shuiwu Zongju Guanyu Jicheng Dianlu Sheji He Ruanjian Chanye Qiye Suodeshui Zhengce De Gonggao (财政部、税务总局关于集成电路设计和软件产业企业所得税政策的公告) [Notice on Enterprise Income Tax Policies for Software and Integrated Circuits Enterprises], Notice of the MoF & STA No. 68 [2019] (promulgated by the MoF & STA on May 17, 2019, effective Jan. 1, 2019); Caizheng Bu, Shuiwu Zongju Guanyu Jicheng Dianlu Shiji Qiye He Ruanjian Qiye 2019 Niandu Qiye Suodeshui Huisuan Qingjiao Shiyong Zhengce De Gonggao (财政部、税务总局关于集成电路设计企业和软件企业2019年度企业所得税汇算清缴适用政策的公告) [Notice on Applicable Policies for Software and Integrated Circuits Enterprises Enterprise Income Tax Payment 2019], Notice of the MoF & STA No. 29 [2020] (promulgated by the MoF & STA on May 29, 2020, effective Jan. 1, 2020).

Software Industry in the New Era, which delineates the parameters for IC products deemed strategically significant to the country, and outlines all the preferential policies they qualify for, spanning from taxation, finance, import and export, education to intellectual property (IP) protection.[158] Following the introduction of the Policies, various ministerial measures were issued in subsequent years, revising earlier versions and specifying new standards under the Policies.[159] According to the current criteria, IC enterprises producing chips of 28-nanometer or more advanced, and which have been operational for over 15 years, are exempted from enterprise income tax (EIT) for ten years.[160] Moreover, R&D expenditures of IC enterprises qualify for a 120 percent EIT deduction.[161] Costs associated with employee training are also deductible from payable EIT, while fixed assets such as equipment, benefit from more preferable depreciation terms.[162] Import tax is either waived or eligible for instalment payment.[163] Furthermore, eligible IC

---

158. Xinshiqi Cujin Jicheng Dianlu Chanye He Ruanjian Qiye Gaozhiliang Fazhan De Ruogan Zhengce (新时期促进集成电路产业和软件产业高质量发展的若干政策) [Several Policies on Promoting the High-Quality Development of Integrated Circuits Industry and Software Industry in the New Era] (promulgated by the State Council on Jul. 27, 2020, effective Jul. 27, 2020).

159. *See, e.g.*, Guanyu Cujin Jicheng Dianlu Chanye He Ruanjian Chanye Gaozhiliang Fazhan Qiye Suodeshui Zhengce De Gonggao (关于促进集成电路产业和软件产业高质量发展企业所得税政策的公告) [Notice on Enterprise Income Tax Policies Promoting the High-Quality Development of Integrated Circuits Industry and Software Industry], Notice of the MoF, NDRC & MIIT No. 45 [2020] (promulgated by the MoF, NDRC & MIIT on Dec. 11, 2020, effective Jan. 1, 2020); Caizheng Bu, Haiguan Zongshu, Shuiwu Zongju Guanyu Zhichi Jicheng Dianlu Chanye He Ruanjian Chanye Fazhan Jinkou Shuishou Zhengce De Tongzhi (财政部、海关总署 税务总局关于支持集成电路产业和软件产业发展进口税收政策的通知) [Notice on Import Tax Policies Supporting the Development of Integrated Circuits Industry and Software Industry], Notice of the MoF, Gen. Admin. of Customs & STA No. 4 [2021] (promulgated by the MoF, Gen. Admin. of Customs & STA on Mar. 16, 2021, effective Jul. 27, 2020); Guanyu Tigao Jicheng Dianlu He Gongye Muji Qiye Yanfa Feiyong Jiaji Kouchu Bili De Gonggao (关于提高集成电路和工业母机企业研发费用加计扣除比例的公告) [Notice on Increasing the R&D Deduction Percentage of Integrated Circuits and Industrial Mother Machine Enterprises], Notice of the MoF, STA, NDRC & MIIT No. 44 [2023] (promulgated by the MoF, STA, NDRC & MIIT on Sept. 12, 2023, effective Jan. 1, 2023).

160. Yiwen Liaojie: *Jicheng Dianlu He Ruanjian Qiye Suodeshui Jianmian Youhui Zhengce (*一文了解：集成电路和软件企业所得税减免优惠政策*)* [*Explainer: Preferential Enterprise Income Tax Deduction and Exemption Policies*], STA (Nov. 12, 2024), https://www.chinatax.gov.cn/chinatax/n810356/n3010387/c5235778/content.html#:~:text=1.%E5%9B%BD%E5%AE%B6%E9%BC%93%E5%8A%B1%E7%9A%84%E9%9B%86,%E5%BE%81%E6%94%B6%E4%BC%81%E4%B8%9A%E6%89%80%E5%BE%97%E7%A8%8E%E3%80%82.

161. Woguo Zhichi Keji Chuangxin Zhuyao Shuifei Youhui Zhengce Zhiyin (我国支持科技创新主要税费优惠政策指引) [Policy Guide on China's Preferential Tax Treatment Supporting Science and Technology Innovation], MoF, Min. of Sci. & Tech. (MST), Gen. Admin. of Customs & STA, https://www.most.gov.cn/kjbgz/202403/W020240313618116200133.pdf, at 123-140.

162. *Id.*

163. *Id.* at 159.

companies enjoy more advantageous input value-added tax deduction policies.[164]

With these policies and initiatives in place, China's domestic chipmaking industry has exhibited remarkable growth, as evidenced by the launches of products such as Huawei Mate 60 Pro and DeepSeek-V3, despite the abovementioned allegations of smuggling Nvidia chips. This growth places U.S. regulators in a challenging predicament: easing export controls could grant China access to cutting-edge technology yet maintaining them might inadvertently fuel China's technological progress.[165] Consequently, the efficacy of these export restrictions remains debatable. However, in August 2025, DeepSeek announced the launch of its latest AI model would be delayed due to a failed attempt to train the model on domestic Huawei chips.[166] It can be observed that the U.S. export controls still have short-term impacts on China's AI capabilities. That said, commentators have likened supply-side control of technology to attempts to "plug holes in a rusty bucket filled with water," which ultimately can only forestall the inevitable technological spread.[167]

---

164. *Id.* at 120.

165. Ariel Cohen, *China's Massive Barrage in the Chip Battle*, FORBES (May 31, 2024), https://www.forbes.com/sites/arielcohen/2024/05/31/chinas-massive-barrage-in-the-chip-battle/.

166. *DeepSeek's Next AI Model Delayed by Attempt to Use Chinese Chips*, FIN. TIMES (Aug. 14, 2025), https://www.ft.com/content/eb984646-6320-4bfe-a78d-a1da2274b092.

167. Daniel H. Joyner, *Economic Nationalism as the Fourth Era of International Trade Law*, 22 MANCHESTER J. INT'L ECON. L. 3, 35 (2025).

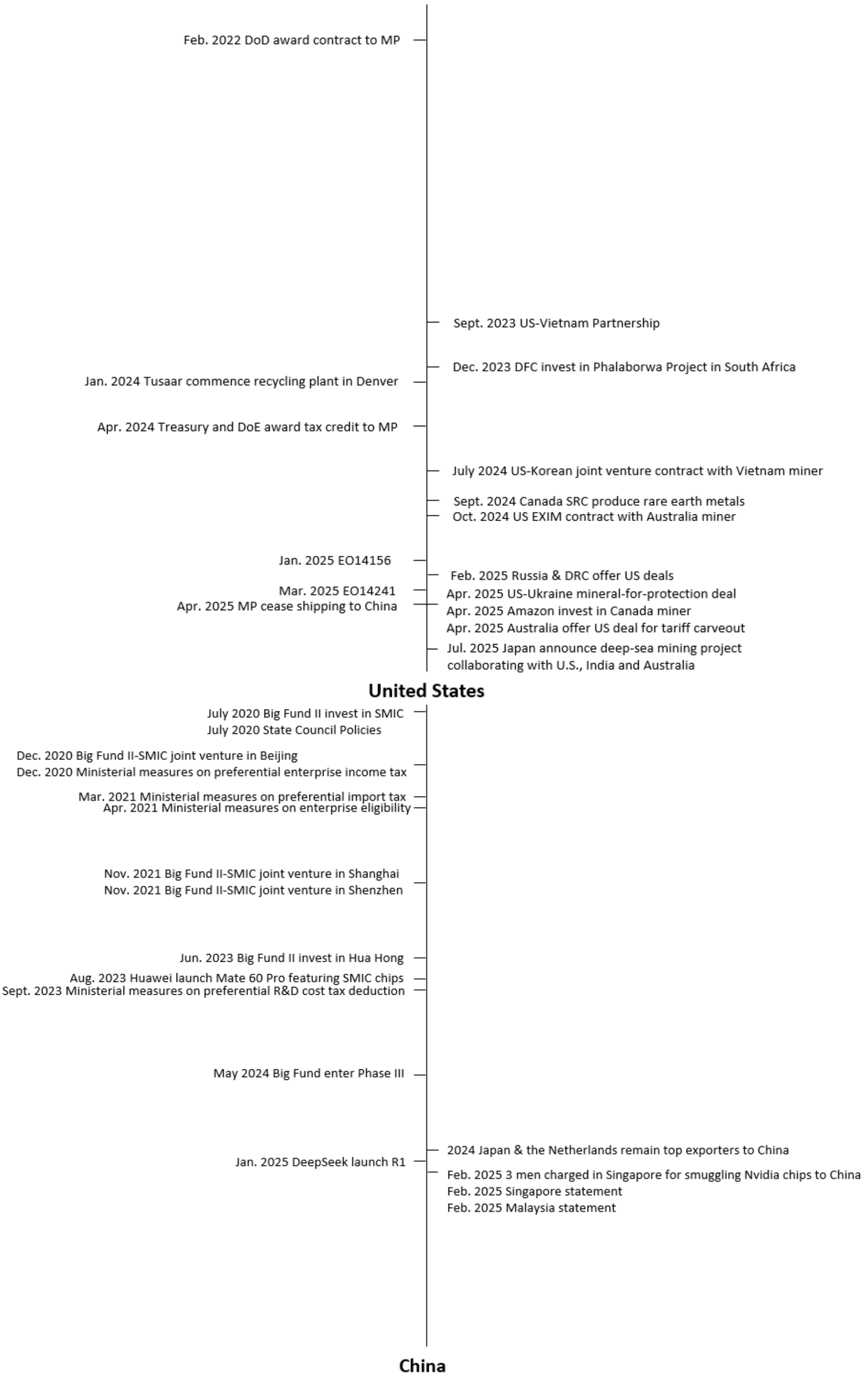

Figure 2 & Figure 3 – Counter-Measures adopted by the United States and China

## IV. Evaluation Under the WTO Framework

Export controls and restrictions traditionally fall under the regulation of the WTO.[168] On December 12, 2022, two months after the United States announced its export controls, China sought consultations at the WTO dispute settlement body (DSB).[169] The subsequent revisions of the request also included the October 2023, April 2024 and December 2024 export controls of the United States.[170] In response, the United States, invoking Article XXI of the GATT, communicated to the DSB that China's request pertained to measures taken to protect U.S. national security, which are political matters not susceptible to review or WTO dispute settlement.[171]

The likelihood of success of this argument hinges on whether semiconductor chips for AI applications fall within the "national security interests" as defined in the GATT. Advocating a narrow interpretation of national security, this article posits that broadening the security exceptions to include semiconductors and AI applications contravenes both the legislative intent of the GATT and previous decisions of the DSB. Moreover, such an expansive application is impractical, as it would almost certainly lead to the strategic overuse of the exception by other nations, ultimately resulting in the dysfunction of the WTO dispute resolution mechanism.

### *A. Whether Security Matters Are Non-Justiciable*

The GATT includes a security exception clause under Article XXI(b). This clause specifies that the Agreement does not prevent any contracting party from taking any action which it considers necessary for the protection of its essential security interests

> (i) relating to fissionable materials or the materials from which they are derived;
>
> (ii) relating to the traffic in arms, ammunition and implements of war and to such traffic in other goods and materials as is carried on directly or indirectly for the purpose of supplying a military establishment;

---

168. *See* CHIEN-HUEI WU, LAW AND POLITICS ON EXPORT RESTRICTIONS (2021).

169. Request for Consultations by China, United States—*Measures on Certain Semiconductor and Other Products, and Related Services and Technologies*, WTO Doc. WT/DS615 (December 12, 2022).

170. Addendum, United States—*Measures on Certain Semiconductor and other Products, and Related Services and Technologies*, WTO. Doc. WT/DS615/1/Rev.1/Add.2 (filed Jan. 10, 2025).

171. Communication from the United States, United States — *Measures on Certain Semiconductor and other Products, and Related Services and Technologies*, WTO. Doc. WT/DS615/4 (filed Jan. 12, 2023).

(iii) taken in time of war or other emergency in international relations.[172]

The United States maintains that this Article authorizes each member state to independently determine which measures are deemed "necessary"' to protect such essential security interests, and asserts that such determinations are not subject to review or capable of resolution by WTO dispute settlement mechanisms.[173] Therefore, the United States did not elaborate on whether or which of the three sub-categories the concerned export controls fall into but simply declared that these measures were "necessary" to protect its "essential security interests."

The pivotal issue in assessing this contention is who decides which measures are "necessary" for the protection of a country's security interests. Specifically, the question hinges on whether the determination of necessity follows an objective (i.e., a measure is objectively necessary according to some pre-set criteria) or subjective standard (i.e., the country implementing the measure deems it necessary). The interpretation by the United States clearly leans toward the latter. However, if this interpretation were to prevail, the inclusion of the three sub-paragraphs describing permissible actions under this clause would be rendered superfluous, as the justifiability of a measure would rely solely on an invoking country's own discretion. Such result would appear inconsistent with the objectives of the WTO and the GATT. On the other hand, a nation does possess inherent sovereignty interests in safeguarding its national security, which is precisely the purpose of this exception clause. Therefore, the latitude granted to member states under the exception is merely a matter of degree.

From a literal interpretation, Article XXI(b) comprises of two distinct parts: First, the adjectival clause "which it considers necessary for the protection of its essential security interests" in the chapeau refers to the preceding word "action," thereby imparting an understanding that member states possess the discretion under Article XXI(b) to determine which "actions" they "consider necessary to protect their essential security interests." Secondly, the three sub-paragraphs, rather than delineating the "essential security interests" that are protectable under this clause, should be interpreted as describing the "actions" that member states are permitted to undertake. This interpretation stems from the text of sub-paragraph (iii), which starts with "taken in time of war…" and therefore clearly pertains to "actions" rather than "interests."[174] Combining the two parts, the actions a party can take under the exception clause must meet two criteria: first, the actions must be considered necessary by the invoking party to protect its essential security interests; secondly, the actions must conform to the description provided in any of the three sub-paragraphs. The determination of the former is subject to a subjective standard, whereas the determination of the latter adheres to an objective standard. [175]

---

172. General Agreement on Tariffs and Trade 1994, Apr. 15, 1994, Marrakesh Agreement Establishing the World Trade Organization, Annex 1A, 1867 U.N.T.S. 187, 33 I.L.M. 1153 (1994) [hereinafter GATT 1994].

173. Communication from the United States, WT/DS615/4, *supra* note 171.

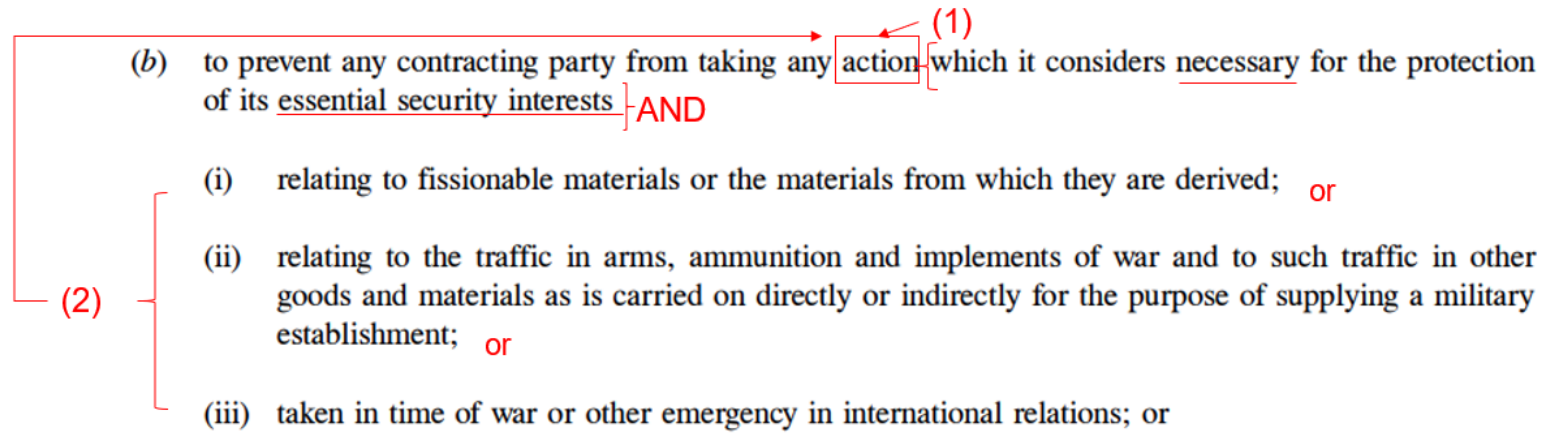

Figure 4 – Inner logic of Article XXI(b)

For the first criterion, the actions must be considered necessary by the invoking party. Although this involves a subjective test, it does not permit a country to arbitrarily declare any measures it sees fit as a security matter based solely on its own discretion. Instead, the country's discretion is constrained by the principle of good faith. In *Russia — Measures Concerning Traffic in Transit*, which marks the first case where a WTO panel examined a party's claims based on the security exception clause, the panel determined that "the obligation of good faith requires that Members not use the exceptions in Article XXI as a means to circumvent their obligations under the GATT 1994."[176] Therefore, the invoking party's definition of "essential security interests" and the nexus between these interests and the measures taken (i.e., necessity) are all governed by the good-faith principle. It falls upon the invoking party to demonstrate that the interest is essential enough to warrant the

---

174. Addendum, *Russia—Measures Concerning Traffic in Transit*, WTO Doc. WT/DS512/R/Add.1 (adopted Apr. 5, 2019), Annex D-5.

175. This understanding is consistent with the view of the WTO panel in *Russia - Measures Concerning Traffic in Transit*, in which the panel found that the phrase "which it considers" covers "necessary" and "its essential security interests." *See* Report of the Panel, *Russia — Measures Concerning Traffic in Transit*, WTO Doc. WT/DS512/R (adopted Apr. 5, 2019).

176. *Id.* at para. 7.133.

implementation of the measure that would otherwise constitute a violation of its GATT obligations.[177]

For the second criterion, an objective standard is applied, which means that the interpretation of the three sub-paragraphs is subject to the objective rules established in international law, rather than the invoking country' own discretion. It is important to note that the principle of good faith governs the interpretation of all aspects of this clause, including the sub-paragraphs. This prevents parties from excessively stretching the explanations to a point where they diverge from the objectives of the GATT.[178] The next part analyzes whether the existing export control measures satisfy the requirements of both two prongs and qualify for a security exception.

### *B. Whether AI-Related Export Controls Qualify for the Security Exceptions*

#### 1. Necessity to Protect Essential Security Interests

As mentioned, an invoking country possesses the discretion to determine whether the export control it implemented is necessary to protect its essential security interests. However, this determination must be made in good faith. This requirement of good faith constrains the invoking country's determination of two key issues: identifying an essential security interest and establishing the necessity of the measures taken.

##### a. Essential Security Interests

Both the United States and China have sought to justify their export control measures on the grounds of national security. In the United States, the ECRA 2018 authorized the BIS to implement a series of export restrictions from 2022 to 2025 for reasons of "national security and foreign policy."[179] Similarly, in China, the authority of the Ministry of Commerce to enact a series of export restrictions stems from its Export Control Law of 2020, which was also promulgated to safeguard "national security and interest."[180] However, the scope and interpretation of national security, as an abstract concept, remain ambiguous and contentious.

---

177. *Id.* at paras. 7.134-7.138.

178. *See* Kentaro Ikeda, *A Proposed Interpretation of GATT Article XXI (b)(ii) in Light of its Implications for Export Control*, 54 CORNELL INT'L L.J. 437, 458–59 (2021).

179. Export Control Reform Act of 2018, H.R. 5040, 115th Cong., (2d Sess. 2018).

180. Export Control Law, *supra* note 59, art. 1.

In international law, interpreting what a term in a treaty denotes shall consult the perspective of an objective third party.[181] The Oxford English Dictionary defines national security as "[t]he safety of a nation and its people, institutions, etc., esp. from *military threat* or from *espionage, terrorism*, etc."[182] While the concept of national security may be considered to encompass various areas of domestic interest, the GATT refines the concept further.[183] "*Essential* security interests", which is evidently a subset of broader security interests, specifically pertain to those aspects critical to a nation's core security. The WTO panel in *Russia* defined the term as "those interests relating to the quintessential functions of the state, namely, the protection of its territory and its population from external threats, and the maintenance of law and public order internally."[184] This definition implies that interests protectable under this clause relate either to external defense (territory and people) or internal sovereignty (law and order).[185] In essence, the "essential security interests" of a country in this context refer to the nation's ability to shield itself and its populace from military invasion and social unrest. For instance, in *Saudi Arabia — Measures Concerning the Protection of Intellectual Property Rights*, the panel recognized Saudi Arabia's interests in safeguarding itself from the "dangers of terrorism and extremism" as related to the "quintessential functions of the state," thus qualifying them as "essential security interests" under the Article.[186]

Indeed, AI is becoming increasingly prevalent in the military domain, given its capabilities to process vast volumes of data efficiently and enhance the autonomy, self-regulation, and self-actuation of combat systems through innate computation and decision-making.[187] For example, AI technologies have facilitated autonomous systems with target detection capabilities, which have been deployed in contemporary battlefields.[188] Other military applications include surveillance, autonomous transportation of weapons, and immersive virtual battle training.[189] The intent of a nation to prevent AI algorithms, specifically designed for such military applications by a rival nation, clearly meets the definition of "essential security interest" under the GATT security exceptions, as they directly pertain to the military capability of the rival nation.

---

181. Kenneth J. Vandevelde, *Treaty Interpretation from a Negotiator's Perspective*, 21 VAND. J. TRANSNAT'L L. 281, 282 (1988).

182. National security, OXFORD ENGLISH DICTIONARY, https://www.oed.com/dictionary/national-security_n?tl=true, (last visited July 4, 2025).

183. Willard L. Thorp, *Trade Barriers and National Security*, 50 AM. ECON. REV. 433, 434 (1960).

184. Report of the Panel, Russia, WT/DS512/R, *supr*a note 175, at para. 7.130.

185. *See, e.g.*, Jovan Babić, *The Military, Law and Sovereignty: Several Remarks about the Military and Its Function in Preserving the Law and State*, 73 VOJNO DELO 133, 141 (2021) (dividing military defense into defense for sovereignty and territorial integrity).

186. Panel Report, *Saudi Arabia — Measures Concerning the Protection of Intellectual Property Rights,* at paras. 7.280-81, WTO Doc. WT/DS567/R (adopted June 16, 2020).

187. Adib Bin Rashid, Ashfakul Karim Kausik, Ahamed Al Hassan Sunny & Mehedy Hassan Bappy, *Artificial Intelligence in the Military: An Overview of the Capabilities*, Applications, and Challenges, INT'L J. INTEL. SYS. 1, 11 (2023).

188. *Id.* at 12–13.

In the October 2022 export control, the BIS noted that the restricted items, including supercomputers, ICs, and semiconductor manufacturing equipment, were used by China in weapon production and military decision-making.[190] Regardless of the necessity issue (i.e., whether the link between the measure and the interests claimed to be protected can be established), which will be discussed in the subsequent section, the stated interests of this batch of export controls were unequivocally defense-focused. However, the subsequent measures of the BIS from 2023 to 2025 have explicitly and repeatedly cited industrial or economic goals, such as "impair[ing] China's development of an indigenous semiconductor ecosystem" or "unlock[ing] unique economic and social benefits," as their objectives.[191] In *Russia*, the panel clarified that the principle of good faith prevented members from re-labelling what are essentially *trade interests*—interests that members had agreed to protect and promote within the WTO system—as *essential security interests*.[192] The panel also observed that, on previous occasions where Article XXI(b)(iii) was invoked to claim an "emergency" or "international crisis," members demonstrated self-restraint to distinguish *military and serious security-related conflicts* from *economic and trade disputes*.[193] Although this observation pertains specifically to the invocation of the third sub-paragraph, it implies that, in the panel's view, security-related conflicts and trade disputes are distinct concepts, being respectively military and economic in nature.

The BIS' export controls were rooted in the U.S. government's strategic pursuit of technological supremacy. Strategic AI policies in the country commenced with the Obama Administration's report on Preparing for the Future of Artificial Intelligence[194] and continued under the Trump Administration's initiative on Maintaining American Leadership in Artificial Intelligence.[195] While the Biden Administration revised the National Artificial Intelligence Research and Development Strategic Plan 2023, building upon its predecessors in 2016 and 2019, to introduce a more principled and reliable AI development regime,[196] the re-elected Trump Administration revoked "harmful" Biden Administration AI policies that "hinder AI innovation" and reverted to the policy goal of promoting American AI leadership.[197] Accordingly, a new police initiative, titled Winning the Race: America's AI Action Plan, was issued on July 23, 2025, directing federal agencies to not only continue to update the existing export controls on semiconductors and fill any loopholes where necessary, but also make efforts to "export" the American export controls to its allies, so that China will not be able to access frontier AI technologies from other countries as well.[198] Meanwhile, efforts are also being made in the legislative branch. In Congress, a bipartisan Global Technology Leadership Act was introduced to the Senate in June 2023, aiming to bolster American competitiveness in crucial emerging technologies, including AI.[199] The necessity of this Act was underscored by the Senators' explicit references to the technological rivalry with China in these strategic domains.[200]

---

189. *Id.* at 17–21.

190. Implementation of Additional Export Controls Order 2022, s*upra* note 1.

191. *See* discussions in Section II.1.

192. Report of the Panel, Russia, WT/DS512/7, *supra* note 175, at para. 7.133.

193. *Id.* at 7.81.

194. EXEC. OFF. OF THE PRESIDENT, NAT'L SCI. & TECH. COUNCIL COMM. ON TECH, PREPARING FOR THE FUTURE OF ARTIFICIAL INTELLIGENCE (2016).

195. Exec. Order No. 13859, 84 Fed. Reg. 3967 (Feb. 11, 2019).

Technological innovation shapes economic power transition.[201] Notably, the friction between the United States and China began in 2016, when presidential candidate Trump criticized the then existing trade policies with China as detrimental to the American economy and American exceptionalism,[202] which subsequently escalated into a trade war between the two countries in 2018. National security was already invoked in 2018 to justify the initiation of the trade war,[203] a move criticized by some scholars as constituting protectionism and "economic nationalism" and not aligning with the historical notion of self-defense.[204] Subsequently,

---

196. SELECT COMM. ON A.I. OF THE NAT'L SCI. & TECH. COUNCIL, NATIONAL ARTIFICIAL INTELLIGENCE RESEARCH AND DEVELOPMENT STRATEGIC PLAN: 2023 UPDATE (2023).

197. THE WHITE HOUSE, FACT SHEET: PRESIDENT DONALD J. TRUMP TAKES ACTION TO ENHANCE AMERICA'S AI LEADERSHIP (2025).

198. THE WHITE HOUSE, WINNING THE RACE: AMERICA'S AI ACTION PLAN (2025) [hereinafter America's AI Plan].

199. Global Technology Leadership Act, S. 1873, 118th Cong. (2023).

200. Michael Bennet, *Bennet, Young, Warner Introduce Bill to Strengthen U.S. Technology Competitiveness*, U.S. SENATORS FOR COLO. (June 8, 2023), https://www.bennet.senate.gov/public/index.cfm/2023/6/bennet-young-warner-introduce-bill-to-strengthen-u-s-technology-competitiveness.

201. *See, e.g.*, JEFFERY DING, TECHNOLOGY AND THE RISE OF GREAT POWERS 3 (2024).

202. *Trump Accuses China of 'Raping' US with Unfair Trade Policy*, BBC (May 2, 2016), https://www.bbc.com/news/election-us-2016-36185012.

203. *See, e.g.*, Jyh-An Lee, *Shifting IP Battlegrounds in the U.S.–China Trade War*, 43 COLUM. J.L. &. ARTS 49, 147, 188-191 (2020).

204. Thomas J. Schoenbaum & Daniel C.K. Chow, *The Perils of Economic Nationalism and A Proposed Pathway to Trade Harmony*, 30 STAN. L. & POL'Y REV. 115 (2019); Chad P. Bown, *Trump's Steel and Aluminum Tariffs Are Cascading out of Control*, PETERSON when trade tension extended to the high-tech sector, although military actors were also involved, it is undeniable that economic and competition considerations played a significant role in shaping these restrictive policies — as expressed in multiple policy documents mentioned above, most of the U.S. export controls had the dual purposes of limiting China's AI capabilities while preserving its own technological leadership. Both dimensions are fundamentally economic in nature.

That said, given the transformative changes ushered in by digital technologies, some scholars have advocated for an expansion of the concept of national security. Traditionally associated with military defense, the various measures adopted by both the United States and China in the broader context of U.S.-China competition have led to a more expansive interpretation of security. The United States views security as retaining superiority in economic, technological, and ideological realms, whereas China considers security to encompass political, social and ideological stability, as well as overseas interests.[205] Under this refined notion of security, a military attack or intrusion is no longer necessary to weaken an adversary, as ideological demoralization, election interference, financial instability, and the destruction of infrastructure could achieve similar effects.[206] Consequently, some may broadly argue that economic and technological powers are crucial as they ultimately bolster military capability,[207] and, conversely, economic damage inflicted on a competitor may translate into a military advantage.[208]

Nevertheless, from an international law perspective, the interests derived from undermining the economy of a competitor to gain a competitive edge are so tenuously connected to the defense and sovereignty interests previously mentioned that they struggle to meet the "essential security interests" test required by the GATT, as exemplified by the core security interests recognized in *Saudi Arabia*.[209] First, to broaden the concept of security interests would contravene the legislative intent of the GATT. Given the GATT's mission to reduce trade barriers and discrimination and taking into account its drafting history, the security exceptions were intended

---

INSTITUTE FOR INT'L ECON. (Feb. 4, 2020), https://www.piie.com/blogs/trade-and-investment-policy-watch/2020/trumps-steel-and-aluminum-tariffs-are-cascading-out.

205. Joel Slawotsky, *Conceptualizing National Security in an Era of Great Power Rivalry: Implications for International Economic Law*, 42 E. ASIA 279 (2025).

206. *Id.*

207. *Id.*

208. Michael Beckley, *Economic Development and Military Effectiveness*, 33 J. STRATEGIC STUD. 43, 54 (2010).

209. Olga Hrynkiv & Lavrijssen Saskia, *Not Trading with the Enemy: The Case of Computer Chips*, 58 J. WORLD TRADE 61, 80 (2024).

to be and shall remain an exceptional clause.[210] However, the current trend of expanding the use of these exceptions indicates an unwanted normalization of what should be an extreme clause. Secondly, from a practical standpoint, broadening the concept of security to encompass economic-centered concerns will almost certainly lead to abuse by member states, as nearly every product could potentially be shielded under such a broad exception rule. Thirdly, from a textual perspective, Article XXI(b) specifies that only *essential* security interests are protected under this clause. Even if countries expanded the concept of national security on their own initiative, these economic-centered security interests can hardly be considered "essential" as defined in the GATT.

Overall, the crux of this debate is that, regardless of whether member states have expanded the scope of national security for domestic law enforcement purposes, these domestic interpretations should not affect the interpretation of Article XXI(b) GATT and the application of international law principles in general. Indeed, establishing an international law governance structure is precisely the purpose of the WTO.[211]

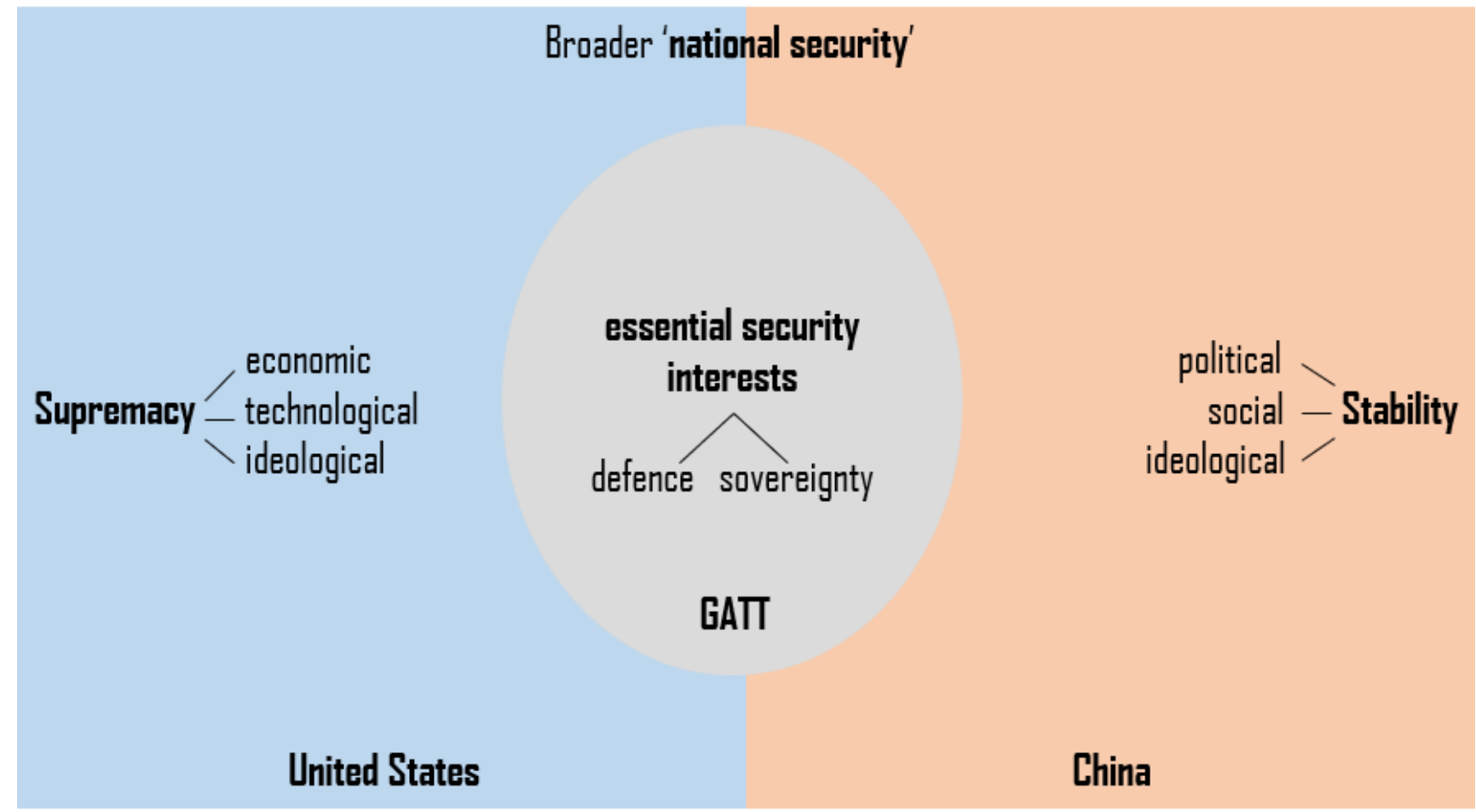

Figure 5 – Relationship between different conceptualizations of security

210. Olga Hrynkiv, *Legal and Policy Reponses to National Security Measures in International Economic Law*, 54 GEO. J. INT'L L. 165, 174, 179 (2023).

211. Pascal Lamy, *The Place of the WTO and Its Law in the International Legal Order*, 17 EUR. J. INT'L L., 969, 972 (2006) ("[T]he WTO rules above all effectively govern the community of its Members … they do form a new legal order as defined above.").

b. Necessity

As mentioned above, the principle of good faith governs not only a country's identification of essential security interests but also its determination of whether a measure is necessary to protect that identified interest. The WTO appellate body in *United States – Import Prohibition of Certain Shrimp and Shrimp Products*, when elucidating Article XXI pertaining to the general exceptions of the GATT, remarked on the principle of good faith as implying that any assertion of a right that affects a field covered by a treaty obligation "must be exercised bona fide, that is to say, *reasonably*."[212] While the subjective standard allows a state to determine for itself which measures "it considers" necessary, such declaration must not be unreasonable from an objective standpoint. Put differently, there must be a nexus between the interests the state proclaims to defend and the measure it declares "necessary."

However, the mere existence of a connection is insufficient. In *Brazil — Measures Affecting Imports of Retreaded Tyres*, also in the context of explaining the general exception clause, the appellate body held that for a measure to be "necessary," a panel must compare the stated objective with the trade restrictiveness of a measure; if a less restrictive alternative exists, the adopted measure cannot be considered truly "necessary."[213] In essence, the measure should be the least restrictive means to achieve the stated interest.[214] Therefore, based on the principle of systematic integration, there is no reason to believe that the necessity requirement in the security exception clause should be construed differently from that in the general exception clause.[215] Thus, Article XXI(b) necessitates a measure to be the *least restrictive* means available to a member to protect its identified essential security interests.

In fact, the United States had commenced its control of semiconductor chips in as early as 2018, with several more export restrictions introduced between 2019 to 2020.[216] Under these earlier controls, an export license could be obtained if the end use was not military.[217] However, the post-2022 bans adopted a much more stringent approach by prohibiting export of dual-use items and reviewing license applications with end users headquartered in China under a presumption-of-denial standard, citing China's "military-civil fusion strategy." These earlier control measures demonstrate that less restrictive measures do exist, and the post-2022 controls covering commercial applications are far from being the *least restrictive* means to safeguard the stated security interests. The BIS has explained that such policy upgrades are intended to close the "loophole" of the previous export control system. However, it was also explicitly stated in the government's remarks that:

> "[g]iven the foundational nature of certain technologies, such as advanced logic and memory chips, we must maintain *as large of a lead as possible*…This has demonstrated that technology export controls can be *more than just a preventative tool*."[218]

---

212. Report of the Appellate Body, *United States — Import Prohibition of Certain Shrimp and Shrimp Products*, WTO Doc. WT/DS58/AB/R (adopted Oct. 12, 1998), citing BIN CHENG, GENERAL PRINCIPLES OF LAW AS APPLIED BY INTERNATIONAL COURTS AND TRIBUNALS ch. 4, para. 125 (1993).

213. Report of the Appellate Body, *Brazil — Measures Affecting Imports of Retreaded Tyres*, WTO Doc. WT/DS332/AB/R (adopted Dec. 3, 2007).

214. Ikeda, *supra* note 178, at 456.

215. Systematic integration is reflected in Article 31(3)(c) of the Vienna Convention. Vienna Convention art. 31(3)(c), May 23, 1969, 1155 U.N.T.S. 331. *See also* Agatha Panday, *The Role of International Human Rights Law in WTO Dispute Settlement*, 16 U.C. DAVIS J. INT'L L. & POL'Y 245, 254 (2009).

216. Hrynkiv & Lavrijssen, *supra* note 209, at 65–67.

It is therefore evident that the recent round of export controls was not intended to be the least restrictive means to protect U.S. national security. As such, these control measures are likely to fail the necessity test under Article XXI(b) GATT.

2. The Three Sub-Paragraphs

In addition to being necessary to protect a recognized essential security interest, the measure must also fit within the description of one of the three sub-paragraphs of Article XXI(b) GATT.

a. Relating to Fissionable Materials or the Materials from which They are Derived

The first sub-paragraph consists of two parts: the measure taken must either relate to fissionable materials *or* to the materials from which fissionable materials are derived. The U.S. DoE defines fissionable materials as "a nuclide capable of sustaining a neutron-induced fission chain reaction," including uranium-233, uranium-235, plutonium-238, plutonium-239, plutonium-241, neptunium-

---

217. *Id.* at 66.

218. Jake Sullivan, *Remarks at the Special Competitive Studies Project Global Emerging Technologies Summit*, THE WHITE HOUSE (Sept. 16, 2022), https://bidenwhitehouse.archives.gov/briefing-room/speeches-remarks/2022/09/16/remarks-by-national-security-advisor-jake-sullivan-at-the-special-competitive-studies-project-global-emerging-technologies-summit/.

237, americium-241, and curium-244.[219] Fissionable materials are derived either from naturally occurring fissile isotopes or from fertile materials through neutron irradiation and nuclear transmutation processes in reactors.[220] These materials are all integral to the production of nuclear power and nuclear weapons. Apparently, neither the United States' export control over semiconductor chips and semiconductor manufacturing equipment nor China's export control over critical minerals and rare earth materials, fall under this category.

b. Relating to the Traffic in Arms, Ammunition and Implements of War and to Such Traffic in Other Goods and Materials as is Carried on Directly or Indirectly for the Purpose of Supplying a Military Establishment

Similarly, sub-paragraph (ii) encompasses two categories of items: (1) those relating to the traffic in arms, ammunition and implements of war, and (2) those relating to the traffic in other goods and materials carried out *directly or indirectly* for the purpose of supplying a military establishment. Most arguments advocating for the expansion of the security exceptions have chosen to analyze semiconductor chips under subparagraph (ii) of the Article.[221]

The first part of the paragraph addresses measures related to traffic[222] in *arms, ammunition* and *implements of war*. Arms refers to "weapons and equipment used to kill and injure people,"[223] and ammunition pertains to "objects that can be shot from a weapon, such as bullets or bombs."[224] These two categories directly relate to military applications. The last category, "implements of war," offers broader coverage than the previous two and thus allows more room for interpretation.[225] The term "implement" originally means "a tool or other piece of equipment for doing work."[226] However, as some scholars have pointed out, interpreting this word too broadly could lead to absurd results, such as classifying the clothes a soldier wears as an "implement of war," thereby potentially shielding cottons, used to produce military garments, under the security exceptions.[227] Given that this term is enumerated alongside arms and ammunition, and considering the inclusion of another catch-all phrase later in the sub-paragraph, "implement of war" should be narrowly interpreted to cover only items primarily used or designed for military purposes,[228] thus excluding clothes, boots or other dual-use products.

---

219. *Fissionable Materials*, U.S. DEP'T OF ENERGY, https://www.directives.doe.gov/terms_definitions/fissionable-materials, (last visited June 30, 2025).

220. THOMAS J. DOLAN, MOLTEN SALT REACTORS AND THORIUM ENERGY 4 (2017); WU YICAN, FUSION NEUTRONICS 73 (2017).

221. *See* Ikeda, *supra* note 178, at 467; Alexandr Svetlicinii & Xueji Su, *The Unsettled Governance of the Dual-Use Items under Article XXI(b)(ii) GATT: A New Battleground for WTO Security Exceptions*, 24 WORLD TRADE REV. 75 (2025); Isabelle Brundieck, *The Validity of Trade Restrictions on Artificial Intelligence Technology Under the General Agreement on Tariffs and Trade's National Security Exception*, 39 AM. U. J. INT'L L. REV. 119, 142.

222. The word "traffic" here may cause confusion as it has multiple meanings that could be applied in this context. However, an investigation into the French version of the GATT will illustrate that the "traffic" here refers to "trade" or "the action of dealing or trading." *See* Ikeda, *supra* note 178, at 465–66.

223. *Arms*, CAMBRIDGE DICTIONARY, https://dictionary.cambridge.org/dictionary/english/arms (last visited July 4, 2025).

224. *Ammunition*, CAMBRIDGE DICTIONARY, https://dictionary.cambridge.org/dictionary/english/ammunition (last visited July 4, 2025).

225. Ikeda, *supra* note 178, at 466.

These excluded items would therefore be considered under the second part of this sub-paragraph. Although designated as a catch-all clause covering "other goods and materials," it stipulates an additional condition: the trading of those goods or material must be for the direct or indirect purpose of supplying a military establishment. This aspect of the security exceptions could, if properly applied, support the United States' end-use and end-user examination procedures prior to 2022 — at that time, a license was available for non-military end uses and end users headquartered in China. If the United States continues to ban the export of semiconductor chips, including those for dual-use purposes, solely to Chinese military or governmental institutions, this transaction could fall within the traffic of *other goods directly* for the purpose of supplying a *military establishment*. If the ban targets some commercial dealers with an anticipated end use or end user being the Chinese military, then the transaction could fall under the traffic of *other goods indirectly* for the purpose of supplying a *military establishment*.

However, the existing export controls, which essentially prohibit every importer headquartered in China from obtaining dual-use semiconductor chips regardless of whether the entity is a commercial or military establishment, or if there is a known relationship between the importer and the Chinese military, exceed the scope of the second half of sub-paragraph (ii). Although the United States has cited China's "military-civil fusion" strategy to justify the expansion of the export controls, the WTO panel in a previous dispute, *United States — Origin Marking Requirement*, has already determined that "the subject matters covered in

---

226. *Implement*, CAMBRIDGE DICTIONARY, https://dictionary.cambridge.org/dictionary/english/implement (last visited July 4, 2025).

227. Ikeda, *supra* note 178, at 466–67.

228. *Id.*

subparagraphs (i) and (ii) [of Article XXI(b)] are clearly related to the defense and military sector."[229]

c. Taken in Time of War or Other Emergency in International Relations

Other scholars analyze semiconductor chips under the third sub-paragraph of Article XXI(b).[230] Notably, this sub-paragraph necessitates recognizing the deterioration of the U.S.-China relation as either a "war" or an "emergency in international relations."

Similar to the logic applied to the previous two sub-paragraphs, the term "war" here should be understood strictly as referring to armed conflicts.[231] Although the competition between the United States and China in the field of AI is often referred to by the media as an "AI war,"[232] such metaphorical language shall not be considered in the legal interpretation of the security exceptions, as a state of war denotes "the ultimate breakdown" in the relations between two countries.[233] Regarding "emergency in international relations", the WTO panel in *United States – Certain Measures on Steel and Aluminium Products* has clarified that this term refers to "situations of a certain gravity or severity and international tensions that are of a critical or serious nature in terms of their impact on the conduct of international relations."[234] The panel in *Russia* construed the inner logic of this sub-paragraph as the first half, "war," being an example of the second half, "emergency in international relations," as indicated by the use of the conjunction "or" and the adjective "other".[235] Thus, it should refer to situations not necessarily amounting to an armed conflict but of the same severity as an armed conflict. In another case, *United States —*

229. Panel Report, *United States – Origin Marking Requirements*, WTO Doc. WT/DS597/R 7.301 (adopted Dec. 21, 2022).

230. *See, e.g.*, Hrynkiv & Lavrijssen, *supra* note 209, at 77–79.

231. Panel Report, *Russia*, WT/DS512/7, *supra* note 175, at para. 7.72; Panel Report, *United States*, WT/DS597/R, *supra* note 229, at para. 7.294.

232. *See, e.g.*, Arjun Kharpal, *U.S.-China Tech Battle Entering Its "Primetime" — and Generative A.I. Could be the Next Frontier*, CNBC (June 22, 2023), http://www.cnbc.com/2023/06/23/us-china-tech-war-why-generative-ai-could-be-the-next-battleground.html?msockid=14788734181869d83c89920119c1682e; Ben Jiang & Bien Perez, *DeepSeek's Tech Breakthrough Hailed in China as Answer to Win AI War*, S. CHINA MORNING POST (Jan. 28, 2025), http://www.scmp.com/tech/tech-trends/article/3296503/deepseeks-tech-breakthrough-hailed-china-answer-win-ai-war; Harold Thibault, *AI Race: US-China Chip War Heats Up*, LE MONDE (May 23, 2025), http://www.lemonde.fr/en/economy/article/2025/05/23/ai-race-us-china-chip-war-heats-up_6741573_19.html.

233. Report of the Panel, *United States*, WT/DS597/R, *supra* note 229, at para. 7.295.

234. Report of the Panel, *United States – Certain Measures on Steel and Aluminium Products*, WTO Doc. WT/DS547/R (adopted Aug. 3, 2023) at para. 7.147.

235. Report of the Panel, *Russia*, WT/DS512/7, *supra* note 175, at paras. 7.71–72.

*Origin Marking Requirement*, the panel, considering the French and Spanish versions of the GATT, held that the seriousness of such an emergency can be best understood as "referring to situations of the *utmost gravity*," which represents a breakdown or near-breakdown of the parties' relations.[236]

Moreover, previous decisions also illuminate the nature of the situations under this sub-paragraph. For example, in *Russia*, the panel referred to the situations elicited in sub-paragraphs (i) and (ii), reasoning that "an emergency in international relations must be understood as eliciting the same type of interests as those arising from the other matters addressed in the enumerated subparagraphs of Article XXI(b)", which are "all defence and military interests, as well as maintenance of law and public order interests."[237] The panel then cited examples covered by the term, which, according to its interpretation, could refer to situations "of latent armed conflict, or of heightened tension or crisis, or of general instability engulfing or surrounding a state," in addition to armed conflicts.[238] Based on the above, the panel concluded that mere political or economic differences between members are insufficient, of themselves, to constitute an "emergency in international relations."[239] Specifically, it mentioned that

> it is normal to expect that Members will, from time to time, encounter political or economic conflicts with other Members or states. While such conflicts could sometimes be considered urgent or serious in a political sense, they will not be 'emergencies in international relations' within the meaning of subparagraph (iii) unless they give rise to defense and military interests, or maintenance of law and public order interests.[240]

As such, the panel's view has been distinctively articulated: political and economic frictions that have not escalated into armed confrontations do not meet the criteria for an "emergency in international relation" as outlined in sub-paragraph (iii) of Article XXI(b) of the GATT. The ongoing dynamics between the United States and China exemplify in this interpretation. Although their relationship is characterized by trade disputes, technological rivalry, and political contention, there has been no descent into armed conflict or disintegration of public order, and efforts towards a diplomatic resolution persist rather than a total complete rupture.

---

236. Report of the Panel, *United States*, WT/DS597/R, *supra* note 229, at para. 7.289, 7.296.

237. Report of the Panel, *Russia*, WT/DS512/7, *supra* note 175, at para. 7.74.

238. *Id.* at para. 7.76.

239. *Id.* at para. 7.75.

240. *Id.*

Moreover, the determination of whether a situation constitutes an emergency is subject to an objective assessment.[241] The panel in *Russia* recognized an emergency in international relations, based on a United Nations General Assembly resolution that acknowledged the presence of armed conflicts between Ukraine and Russia.[242] Similarly, in *Saudi Arabia*, an emergency was declared due to the complete breakdown of Saudi Arabia's diplomatic and economic relations with Qatar.[243] However, in *United States — Origin Marking Requirement*, the panel did not accept that the human rights issues in Hong Kong reached the severity necessary to be deemed an "emergency in international relations."[244] Indeed, the panel in that case pointed out that, despite the myriad of political, economic, social, and environmental pressures globally, contracting members generally continue to navigate their international relations within a range of international legal frameworks designed to maintain predictability and stability within the global system.[245] Therefore, the current U.S.-China situation is unlikely to be deemed an "emergency" that could justify the extensive export controls on semiconductor chips.[246]

In conclusion, these post-2022 export restrictions do not concern an essential security interest under a stringent interpretation of the term. Even if such an interest were considered "essential", the expansive nature of the controls would likely fail the necessity test for not being the least restrictive means available. Furthermore, the export controls do not align with any of the three sub-paragraphs, all of which necessitate a scenario akin to an armed conflict.

### *C. Whether the Alleged Violations are Likely to Prevail, and Beyond…*

Lastly, if the measures are not encompassed by the security exceptions, the matter will then proceed to an assessment of the merits of China's claims regarding the United States' violation of its trade obligations. Among these allegations are violations related to the most-favored-nation treatment and the elimination of quantitative restrictions.[247] While the likelihood of these allegations succeeding is beyond the scope of this study, it is important to note that even if a ruling favorable to China is issued, the panel report cannot be enforced should the United States choose to appeal to the non-functional WTO appellate body, which has been inoperative since December 2019.[248]

---

241. *Id.* at paras. 7.76–77.
242. *Id.* at paras. 7.122–23.
243. Report of the Panel, *Saudi Arabia*, WT/DS567/R, *supra* note 186.
244. Report of the Panel, *United States*, WT/DS597/R, *supra* note 229, at para. 7.358.
245. *Id.* at para. 7.311.
246. *See, e.g.*, Joyner, *supra* note 167, at 35.

In response, China has implemented its own retaliatory export controls without awaiting a final determination from the WTO panel.[249] If challenged, China's export controls are also likely to fail the requirements of Article XXI(b), following the same set of criteria. However, such retaliatory measures outside the WTO framework may be the only feasible option available under the prevailing circumstances.[250] The simultaneous initiation of a WTO procedure and retaliation underscores member states' lack of confidence in the WTO dispute resolution mechanism, and the strategic position of these procedures as a leverage for bilateral negotiations outside the WTO framework.[251] During a press conference held on June 27, 2025, the Chinese Ministry of Commerce confirmed that, following the high-level talks in London, a consensus was reached whereby the United States would remove a series of restrictive measures previously imposed on China, in exchange for China authorizing the export of certain regulated products.[252] Similarly, U.S. Treasury Secretary Scott Bessent indicated that the two countries have resolved issues concerning the shipments of rare earth minerals and magnets to the United States, signaling a "de-escalation" of the export controls. [253]

This result suggests a consensus that bilateral dialogue has proven a more effective method for addressing export control issues than the traditional WTO/GATT framework. However, despite this statement, the recently released America's AI Action has pointed to the opposite direction. Stating that "America must impose strong export controls on sensitive technologies…[and] encourage partners and allies to follow U.S. controls, and not backfill," the initiative made it imperative to expand the existing controls and develop new controls in coordination with U.S. allies, and signaled the possibility of using the Foreign Direct Product Rule and secondary tariffs to achieve this international alignment.[254] The pessimistic view is that the excessive use of export controls, as one of the numerous strategies adopted by nations to promote "economic nationalism", demonstrates that the subjective national security discretion, driven by political and economic factors, is fundamentally incompatible with the objective international trade rules designed to eliminate trade discrimination.[255] Taken under a greater picture, these export controls have contributed to and is likely to further enhance the current geopolitical fragmentation.

---

247. GATT 1994, *supra* note 172, arts. I:1. XI:1.

248. Isabelle van Damme, *25 Years of Law and Practice at the WTO: Did the Appellate Body Dig Its Own Grave?*, 26 J. INT'L ECON. L. 124, 124 (2023). The United States is not a member to the Multi-Party Interim Appeal Arbitration Arrangement (MPIA), which was a separate appeal system set up by 16 WTO members.

249. Hrynkiv & Lavrijssen, *supra* note 209, at 83.

250. *Id.* at 82.

251. *Id.* at 84.

252. *Shangwu Bu Xinwen Fayanren Jiu Zhongmei Lundun Kuangjia Youguan Qingkuang Da Jizhe Wen* (商务部新闻发言人就中美伦敦框架有关情况答记者问) [*Ministry of Commerce Spokesperson Answers Questions regarding China-U.S. London Framework at Press Conference*], MIN. OF COM. NEWS OFF. (June 27, 2025), https://www.mofcom.gov.cn/xwfb/xwfyrth/art/2025/art_86bfd1f5c4a34e4c91bff252c50a0cbc.html.

253. Elaine Kurtenbach & Will Weissert, *Bessent Says New U.S.-China Deal is a Sign of 'De-Escalation' After Trump's Trade War Roiled Global Markets*, FORTUNE (June 27, 2025), http://www.fortune.com/2025/06/27/bessent-china-trump-trade-deal-tariffs-rare-earth-minerals/.

## V. CONCLUSION

Recognizing the transformative influence of frontier technologies, the battle for economic supremacy between the United States and China has extended into the technological arena, particularly concerning AI. The intensifying rivalry in the AI sector has spurred both nations to maximize their own AI capabilities while simultaneously attempting to stifle the AI advancements of their competitor. Against this backdrop, export control measures have been strategically employed by both countries to hinder each other from achieving breakthroughs in AI technology.

Considering their respective strength and vulnerabilities, the United States has imposed restrictions on the export of semiconductor chips and chipmaking equipment, whereas China has focused its export controls on critical minerals and REEs. Although these measures may inflict short-term setbacks to their adversaries, their impact remains relatively limited due to both nations' capacity to source these restricted items from alternative suppliers. Moreover, such export controls may unintentionally encourage domestic R&D in the restricted technologies within the rival nation. The balance between the short-term impediments and the potential long-term benefit remains debatable.

These measures have also led to increasing trade frictions and disputes under international law. While countries often justify their actions on national security grounds, this article advocates for a narrower interpretation of security interests that would qualify for an Article XXI(b) exception under the GATT. Under this constrained interpretation, economic security does not fall under the "essential security interests" recognized by the GATT, and export controls driven by economic motives fail to meet the necessity test and do not align with any of the three sub-paragraphs of the Article. Indeed, the security exceptions were intended for the application in extreme situations, where political and economic tensions between members are commonplace, with the expectation that members will collaborate amicably to resolve such issues. In this light, although the WTO proceeding concerning these export control measures between the two countries is still underway, it is more probable that the dispute will be resolved through bilateral negotiations outside the WTO framework.

---

254. America's AI Plan, *supra* note 198, Pillar III.
255. Joyner, *supra* note 167, at 40.